\documentclass[twocolumn]{aastex61}
\usepackage{datetime} 

\newcommand{\titlerunning}[1]{\shorttitle{#1}}
\newcommand{\authorrunning}[1]{\shortauthors{#1}}

\newcommand*\inst[1]{\unskip\hbox{\@textsuperscript{\normalfont$#1$}}}

\newcommand*\institute[1]{
  \begingroup
    \let\and\relax
    \renewcommand*\inst[1]{}%
    \renewcommand*\thanks[1]{}%
    \renewcommand*\email[1]{}%
  \endgroup
  \newcommand{\institutions}{#1}
}%

\let\oldarcsec\arcsec
\renewcommand\arcsec{\oldarcsec\xspace}%

\renewcommand{\ion}[2]{\textup{#1\,\textsc{\lowercase{#2}}}}

\usepackage{latexsym}		
\usepackage{graphicx}		
\usepackage{rotating}		
\usepackage{natbib}  
\usepackage{savesym}
\usepackage{amssymb}
\usepackage{amsmath}
\usepackage{morefloats}
\savesymbol{doublespace}
\usepackage{xspace}
\usepackage{color}
\usepackage{mdframed}
\usepackage{url}
\usepackage{subfigure}
\usepackage{grffile}
\usepackage{import}
\usepackage[utf8]{inputenc}
\usepackage{booktabs}
\usepackage{fancyhdr}

\usepackage[hang,flushmargin]{footmisc}
\usepackage{ifpdf}

\newcommand{\msun}{\ensuremath{M_{\odot}}\xspace}			

\newcommand{\lsun}{\ensuremath{L_{\odot}}\xspace}			
\newcommand{\hh}{\ensuremath{\textrm{H}_{2}}\xspace}			

\newcommand{\methanol}{\ensuremath{\textrm{CH}_3\textrm{OH}}\xspace}

\newcommand{\cyanoacetylene}{\ensuremath{\textrm{HC}_{3}\textrm{N}}\xspace}

\newcommand{\water}{H$_{2}$O\xspace}		

\newcommand{\hii}{\ion{H}{ii}\xspace}

\newcommand{\kms}{\textrm{km~s}\ensuremath{^{-1}}\xspace}	

\newcommand{\permyr}{\ensuremath{\mathrm{Myr}^{-1}}\xspace}
\newcommand{\pers}{\ensuremath{\mathrm{s}^{-1}}\xspace}
\newcommand{\perspc}{\ensuremath{\mathrm{pc}^{-2}}\xspace}

\newcommand{\percc}{\ensuremath{\textrm{cm}^{-3}}\xspace}

\newcommand{\persc}{\ensuremath{\textrm{cm}^{-2}}\xspace}

\newcommand{\peryr}{\ensuremath{\textrm{yr}^{-1}}\xspace}
\newcommand{\um}{\ensuremath{\mu \textrm{m}}\xspace}    

\newcommand{\ammonia}{NH\ensuremath{_3}\xspace}

\def\ee#1{\ensuremath{\times10^{#1}}}

\newcommand{\perbeam}{\ensuremath{\textrm{beam}^{-1}}\xspace}

\def\eqref#1{Equation \ref{#1}}

\def\Figure#1#2#3#4#5{
\begin{figure*}[!htp]
\includegraphics[scale=#4,width=#5]{#1}
\caption{#2}
\label{#3}
\end{figure*}
}

\def\FigureOneCol#1#2#3#4#5{
\begin{figure}[!htp]
\includegraphics[scale=#4,width=#5]{#1}
\caption{#2}
\label{#3}
\end{figure}
}

\def\FigureTwo#1#2#3#4#5#6{
\begin{figure*}[!htp]
\subfigure[]{ \includegraphics[scale=#5,width=#6]{#1} }
\subfigure[]{ \includegraphics[scale=#5,width=#6]{#2} }
\caption{#3}
\label{#4}
\end{figure*}
}

\newenvironment{rotatepage}
{}{}

\newcommand{\githash}{e26ce0f}\newcommand{\gitdate}{2017-12-28\xspace}\newcommand{\ncores}{271\xspace}
\newcommand{\nhii}{31\xspace}
\newcommand{\nmasermatch}{10\xspace}
\newcommand{\nxraymatch}{3\xspace}

\newcommand{\sfrdynagemax}{0.062\xspace}
\newcommand{\sfrbrickagemax}{0.11\xspace}
\newcommand{\nalphas}{62\xspace}
\newcommand{\ngttwo}{33\xspace}
\newcommand{\dsgrb}{\ensuremath{8.4\textrm{~kpc}}\xspace}
\newcommand{\percent}{\%\xspace}

\begin{document}
\title{Distributed star formation throughout the Galactic Center cloud Sgr B2}
\titlerunning{Sgr B2 ALMA}
\authorrunning{Ginsburg et al}
\newcommand{\nraojansky}{\affiliation{\it{Jansky fellow of the National Radio Astronomy Observatory, 1003 Lopezville Rd, Socorro, NM 87801 USA }}}
\newcommand{\nrao}{\affiliation{\it{National Radio Astronomy Observatory, 1003 Lopezville Rd, Socorro, NM 87801 USA }}}
\newcommand{\nraocv}{\affiliation{\it{National Radio Astronomy Observatory, 520 Edgemont Rd, Charlottesville, VA 22903, USA }}}
\newcommand{\eso}{ \affiliation{\it{ European Southern Observatory, Karl-Schwarzschild-Stra{\ss}e 2, D-85748 Garching bei München, Germany } } }

\newcommand{\radboud}{\affiliation{\it{Department of Astrophysics/IMAPP, Radboud University Nijmegen, PO Box 9010, 6500 GL Nijmegen, the Netherlands}}}
\newcommand{\allegro}{\affiliation{\it{ALLEGRO/Leiden Observatory, Leiden University, PO Box 9513, 2300 RA Leiden, the Netherlands}}}
\newcommand{\zah}{\affiliation{\it{Astronomisches Rechen-Institut, Zentrum f{\"u}r Astronomie der Universit{\"a}t Heidelberg, M{\"o}nchhofstra{\ss}e 12-14, 69120 Heidelberg, Germany}}}
\newcommand{\casa}{\affiliation{\it{CASA, University of Colorado, 389-UCB, Boulder, CO 80309}} }
\newcommand{\jodrell}{\affiliation{\it{Jodrell Bank Centre for Astrophysics, School of Physics and Astronomy, University of Manchester, Oxford Road, Manchester M13 9PL, UK}}}
\newcommand{\morelia}{\affiliation{\it{Instituto de Radioastronom{\'i}a y Astrof{\'i}sica, UNAM, A.P. 3-72, Xangari, Morelia, 58089, Mexico}}}
\newcommand{\sjsu}{\affiliation{\it{{San Jose State University, One Washington Square, San Jose, CA 95192}}}}
\newcommand{\herts}{\affiliation{\it{Centre for Astrophysics Research, University of Hertfordshire, College Lane, Hatfield, AL10 9AB, UK}}}
\newcommand{\uofa}{\affiliation{\it{Dept. of Physics, University of Alberta, Edmonton, Alberta, Canada}}}
\newcommand{\arcetri}{\affiliation{\it{INAF-Osservatorio Astrofisico di Arcetri, Largo E. Fermi 5, I-50125, Florence, Italy } } }
\newcommand{\exclus}{\affiliation{\it{Excellence Cluster Universe, Boltzman str. 2, D-85748 Garching bei M\"unchen, Germany } }}
\newcommand{\ljmu}{\affiliation{\it{Astrophysics Research Institute, Liverpool John Moores University, 146 Brownlow Hill, Liverpool L3 5RF, UK }}}
\newcommand{\koeln}{\affiliation{\it{I. Physikalisches Institut, Universi\"at zu K\"oln, Z\"ulpicher Str.\ 77, 50937 K\"oln, Germany}}}
\newcommand{\mpia}{\affiliation{\it{Max-Planck-Institute for Astronomy, Koenigstuhl 17, 69117 Heidelberg, Germany}}}
\newcommand{\agnesscott}{\affiliation{\it{Agnes Scott College, 141 E. College Ave., Decatur, GA 30030}}}
\newcommand{\chile}{\affiliation{\it{Departamento de Astronom{\'i}a, Universidad de Chile, Casilla 36-D, Santiago, Chile}}}
\newcommand{\leiden}{\affiliation{\it{Leiden Observatory, Leiden University, PO Box 9513, NL-2300 RA Leiden, the Netherlands }}}
\newcommand{\mpe}{\affiliation{\it{Max-Planck-Institut für extraterrestrische Physik, D-85748 Garching, Germany}}}
\newcommand{\boston}{\affiliation{\it{Boston University Astronomy Department, 725 Commonwealth Avenue, Boston, MA 02215, USA}}}
\newcommand{\cfa}{\affiliation{\it{Harvard-Smithsonian Center for Astrophysics, 60 Garden St. Cambridge, MA 02138}}}
\newcommand{\usf}{\affiliation{\it{University of South Florida, Physics Department, 4202 East Fowler Ave, ISA 2019 Tampa, FL 33620}}}
\newcommand{\uconn}{\affiliation{\it{University of Connecticut, Department of Physics, 2152 Hillside Rd., Storrs, CT 06269}}}

\newcommand{\jao}{\affiliation{\it{Joint ALMA Observatory, Alonso de Córdova 3107, Vitacura, Santiago, Chile}}}
\newcommand{\naoj}{\affiliation{\it{National Astronomical Observatory of Japan, Alonso de Córdova 3788, 61B Vitacura, Santiago, Chile}}}
\newcommand{\naojtwo}{\affiliation{\it{National Astronomical Observatory of Japan, 2-21-1 Osawa, Mitaka,Tokyo, 181-8588, Japan}}}

\author[0000-0001-6431-9633]{Adam Ginsburg}
\nraojansky
\eso

\correspondingauthor{Adam Ginsburg}
\email{aginsbur@nrao.edu; adam.g.ginsburg@gmail.com}

\author{John Bally}
\casa

\author{Ashley Barnes}
\ljmu

\author{Nate Bastian }
\ljmu

\author{Cara Battersby}
\cfa
\uconn

\author{Henrik Beuther}
\mpia

\author{Crystal Brogan}
\nraocv

\author{Yanett Contreras}
\leiden

\author{Joanna Corby}
\nraocv
\usf

\author{Jeremy Darling}
\casa

\author{Chris De~Pree}
\agnesscott

\author{Roberto Galv{\'a}n-Madrid}
\morelia

\author{Guido Garay}
\chile

\author{Jonathan Henshaw}
\mpia

\author{Todd Hunter}
\nraocv

\author{J.~M.~Diederik Kruijssen}
\zah

\author{Steven Longmore}
\ljmu

\author{Xing Lu}
\naojtwo

\author{Fanyi Meng}
\koeln

\author{Elisabeth A.C. Mills }
\sjsu
\boston

\author{Juergen Ott}
\nrao

\author{Jaime E. Pineda}
\mpe

\author{{\'A}lvaro S{\'a}nchez-Monge}
\koeln
\author{Peter Schilke}
\koeln
\author{Anika Schmiedeke}
\koeln
\mpe

\author{Daniel Walker}
\ljmu
\jao
\naoj

\author{David Wilner}
\cfa
\begin{abstract}
We report ALMA observations with resolution $\approx0.5$\arcsec at 3 mm  of the
extended Sgr B2 cloud in the Central Molecular Zone (CMZ). We detect \ncores
compact sources, most of which are smaller than 5000 AU.  By ruling out
alternative possibilities, we conclude that these sources consist of a mix of
hypercompact \hii regions and young stellar objects (YSOs).  Most of the
newly-detected sources are YSOs with gas envelopes which, based on their
luminosities, must contain objects with stellar masses $M_*\gtrsim8$ \msun.
Their spatial distribution spread over a $\sim12\times3$ pc region demonstrates
that Sgr B2 is experiencing an extended star formation event, not just an
isolated `starburst' within the protocluster regions.  Using this new sample,
we examine star formation thresholds and surface density relations in Sgr B2.
While all of the YSOs reside in regions of high column density
($N(\hh)\gtrsim2\ee{23}$ \persc), not all regions of high column density
contain YSOs.  The observed column density threshold for star formation is
substantially higher than that in solar vicinity clouds, implying either that
high-mass star formation requires a higher column density or that any star
formation threshold in the CMZ must be higher than in nearby clouds.  The
relation between the surface density of gas and stars is incompatible with
extrapolations from local clouds, and instead stellar densities in Sgr B2
follow a linear $\Sigma_*-\Sigma_{gas}$ relation, shallower than that observed
in local clouds.  Together, these points suggest that a higher volume density
threshold is required to explain star formation in CMZ clouds.

\end{abstract}

\section{Introduction}

\Figure{f1}
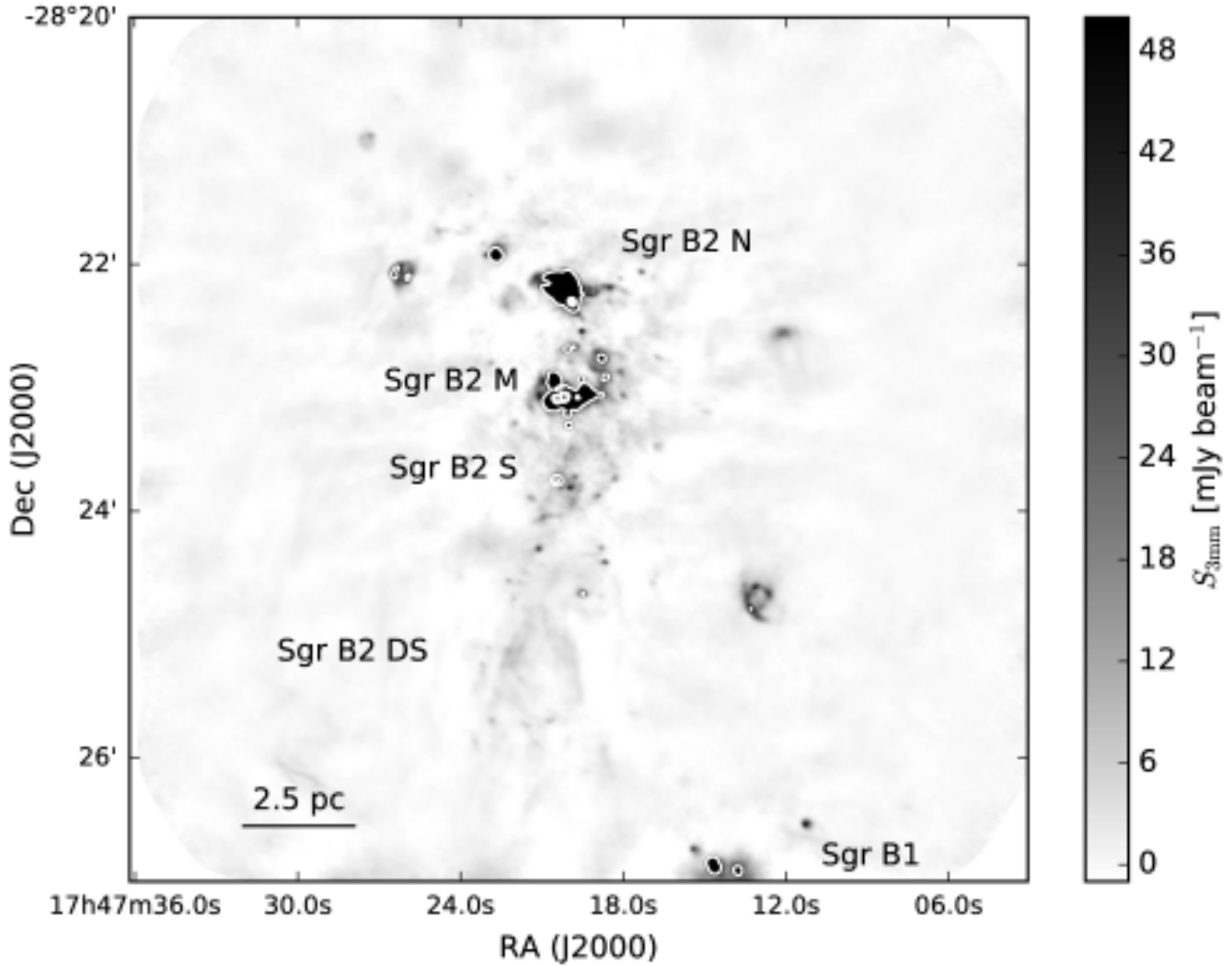
{An overview of the Sgr B2 region, with the most prominent regions labeled.
The image shows the ALMA 3 mm observations imaged with 1.5\arcsec resolution
to emphasize the larger scale emission features.  White contours
are included at [50, 500, 1000, 1500, 2000] mJy/beam to show the flux levels
of the saturated regions.  For a cartoon version of this
figure, see \citet{Schmiedeke2016a} Figure 1.}
{fig:overview}{1}{\textwidth}

The Central Molecular Zone (CMZ) of our Galaxy appears to be overall deficient
in star formation relative to the gas mass it contains
\citep{Guesten1983b,Morris1996a,Beuther2012a,Immer2012a,Longmore2013b,
Kauffmann2017c,Kauffmann2017b,Barnes2017b}.  This deficiency
suggests that star formation laws, i.e., the empirical relations between
the star formation rate and gas surface density, are not universal.  The gas
conditions in the Galactic center are different from those in nearby clouds,
providing a long lever arm in a few
parameters \citep[e.g., pressure, temperature, velocity
dispersion;
][]{Kruijssen2013a,Ginsburg2016a,Immer2016a,Shetty2012a,Henshaw2016a} that
facilitates measurements of the influence of environmental effects on star
formation.

The CMZ dust ridge contains most of the dense molecular material in the
Galactic center \citep{Lis1999a,Bally2010a,Molinari2011a}.  The observed star
formation deficiency comes from comparing the quantity of dense gas to  star
formation tracers such as water masers and free-free emission
\citep{Longmore2013b}, infrared source counts \citep{Yusef-Zadeh2009a}, or
integrated infrared luminosity \citep{Barnes2017b}. 

Recent searches for ongoing star formation using high-resolution millimeter
observations of selected clouds in the CMZ have revealed few star-forming cores
\citep{Johnston2014a,Rathborne2014a,Rathborne2015a,Kauffmann2017c,Kauffmann2017b}.
As summarized by \citet{Barnes2017b}, most of the dust ridge clouds contain
$<1000$ \msun of stars, or $\sim2\%$ of their mass in stars.  The Sgr B2 N
(North), M (Main), and S (South) protoclusters \citep[][Figure
\ref{fig:overview}]{Schmiedeke2016a} are exceptional in that they  are actively
forming star clusters and contain high-mass YSOs (young stellar objects) and
many compact \hii regions \citep[e.g.,][]{Higuchi2015a,Gaume1995a}; despite the
active star formation, the overall cloud appears to be as inefficient as the
other dust ridge clouds \citep{Barnes2017b}.  Besides Sgr B2, a few of the dust
ridge regions are forming stars at a much lower level, including the 20 \kms
and 50 \kms clouds \citep{Lu2015b,Lu2017a}, Sgr C \citep{Kendrew2013a}, and
dust ridge Clouds C, D, and E \citep[Walker et al, in
prep;][]{Ginsburg2015b,Barnes2017b}.  These regions contain only a small number
of high-mass cores, YSOs, and small \hii regions.

Most observations of the Sgr B2 cloud focus on the ``hot cores'' Sgr B2 N and
M, which are high-mass protoclusters (they are likely to form clusters with
$M\gtrsim10^4$ \msun).  The extended cloud has been the subject
of some studies in gas tracers, but it has never been observed at high
($\lesssim10\arcsec$) resolution in the far infrared or millimeter regime.
Radio observations at $\nu<25$ GHz have revealed extended \ammonia and several
masers \citep{Martin-Pintado1999a,McGrath2004a,Caswell2010a}, but these tracers
only detect a subset of star-forming sources.  \citet{Martin-Pintado1999a}
suggested the presence of ongoing star formation in the broader Sgr B2 cloud
based on the detection of three \ammonia (4,4) `hot cores' south of Sgr B2 S.
Despite this suggestion, and the high density of gas throughout the broader Sgr
B2 cloud, an extended star formation event has not been verified.

We report the first observations of extended, ongoing star formation in the Sgr
B2 cloud.  We observed a $\sim15\times15$ pc section of the Sgr B2 cloud and
identified star formation along the entire molecular dust ridge known as Sgr B2
Deep South \citep[DS, also known as the `Southern
Complex';][]{Jones2012a,Schmiedeke2016a}.  These observations allow us to
perform one of the best star-counting based determinations of the star
formation rate within the dense molecular gas of the CMZ.

We adopt a distance to Sgr B2 $D_{\mathrm{Sgr B2}}=\dsgrb$, which is consistent
with Sgr B2 being located in the CMZ dust ridge.  While \citet{Reid2009a}
measure a closer distance of $7.9\pm0.8$ kpc, and \citet{Boehle2016a}
measure a distance to Sgr A$^*$ $7.86\pm0.14$ kpc, we use a value closer to the
IAU-recommended Galactic Center distance of 8.5 kpc, accounting for the
distance difference of $\approx100$ pc measured by
\citet{Reid2009a}\footnote{\citet{Reid2014a} also conclude that the distance to
the Galactic center is 8.34 kpc, suggesting that the direct parallax
measurement to Sgr B2 is underestimated.}.  Choosing the closer distance would
result in masses and luminosities smaller by 12\%, which would not affect any
of the conclusions of this paper.

We describe the new ALMA observations and the archival single-dish data used
to estimate gas column density in Section
\ref{sec:observations}. We focus on the continuum sources selected from the
ALMA data, which we identify in Section \ref{sec:contsources}.  In Section
\ref{sec:analysis}, we perform catalog cross-matching (\S
\ref{sec:crossmatch}) and classify the sources (\S
\ref{sec:classification}).  In Section \ref{sec:sfdiscussion}, we discuss the
star formation rate and flux distribution (\S \ref{sec:distributionsandsfr}),
the
relation between the clusters and the extended star forming population (\S
\ref{sec:clustersandextended}), some implications of our observations
for turbulent star formation theories (\S \ref{sec:turbsftheory}),
and examine star formation thresholds (\S \ref{sec:thresholds})
and surface density relations (\S \ref{sec:gutermuth}).   We conclude
in Section \ref{sec:conclusions}.  Afterward, several appendices describe the
single-dish combination (Appendix \ref{sec:singledishcomb}), self-calibration
(Appendix \ref{sec:selfcal}), and the photometric catalog (Appendix
\ref{sec:catalog}).  Three more appendices show additional figures of
\cyanoacetylene (Appendix \ref{sec:hc3nfigures}), archival VLA 1.3 cm continuum
data (Appendix \ref{sec:onept3cm}), and an additional comparison of the surface
density relations to other works (Appendix
\ref{sec:ladasurfdensappendix}).

\FigureTwo
{f2}
{f3}
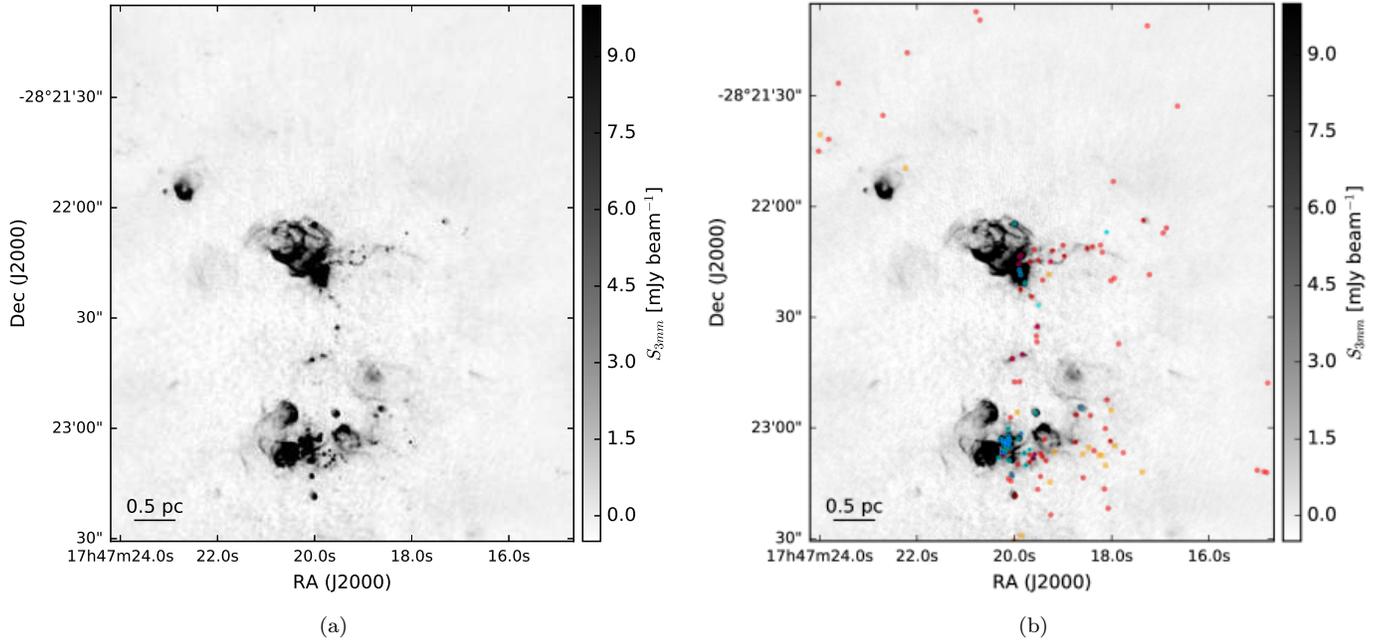
{Images of the ALMA 3 mm continuum in the Sgr B2 M and N region.  The right
figure additionally includes markers at the position of each identified
continuum pointlike source: red dots are `conservative', high-confidence
sources, orange squares are `optimistic', low-confidence sources, cyan are \hii
regions, magenta +'s are \methanol masers, blue +'s are \water masers, and
green X's are X-ray sources.  The massive
protocluster Sgr B2 M is the collection of \hii regions and compact sources in
the lower half of the image.  The other massive protocluster, Sgr B2 N, is in
the center.
The crowded parts of the images are shown with inset zoom-in panels in
Figure \ref{fig:MandNzooms}.
}
{fig:continuumMandN}{1}{0.5\textwidth}

\section{Observations and Data Reduction}
\label{sec:observations}
\subsection{ALMA data}
Data were acquired as part of ALMA project 2013.1.00269.S.  Observations were
taken in ALMA Band 3 with the 12m Total Power array, the ALMA ACA 7m array, and
in two configurations with the ALMA 12m array; durations and dates of the
observations and details of the array configurations are listed in Table
\ref{tab:observations}.  The setup included the maximum allowed number of
channels, 30720, across 4 spectral windows in a single polarization; the
single-polarization mode was adopted to support moderate spectral resolution
($\sim0.8$ \kms, 244 kHz channels) across the broad bandwidth.  The basebands
were centered at 89.48, 91.28, 101.37, and 103.23 GHz with bandwidth 1.875 GHz
(total 7.5 GHz).  The off position used to calibrate the system temperature for
the Total Power (TP) observations was at J2000 17:52:06.461
-28:30:32.095.

\begin{table*}[htp]
\centering
\caption{Observation Summary}
\begin{tabular}{lllll}
\label{tab:observations}
Date & Array & Observation Duration &  Baseline Length Range  & \# of antennae\\
     &       & seconds              & meters                    & \\
\hline
01-Jul-2014 & 7m & 4045 & 9-49 & 10\\
02-Jul-2014 & 7m & 4043 & 9-49 & 10\\
03-Jul-2014 & 7m & 7345 & 9-48 & 8\\
06-Dec-2014 & 12m & 6849 & 15-349 & 34\\
01-Apr-2015 & 12m & 3464 & 15-328 & 28\\
02-Apr-2015 & 12m & 3517 & 15-328 & 39\\
01-Jul-2015 & 12m & 3517 & 43-1574 & 43\\
02-Jul-2015 & 12m & 10598 & 43-1574 & 42\\

\hline

25-Jan-2015 & TP & 6924  & - & 3\\
01-Apr-2015 & TP & 1986  & - & 2\\
11-Apr-2015 & TP & 6920  & - & 3\\
12-Apr-2015 & TP & 10441 & - & 3\\
25-Apr-2015 & TP & 13928 & - & 3\\
26-Apr-2015 & TP & 22562 & - & 3\\
18-May-2015 & TP & 8342  & - & 3\\
\hline
\end{tabular}
\end{table*}

The ALMA QA2 calibrated measurement sets were combined to make a single
high-resolution, high-dynamic range data set.  We imaged the continuum jointly
across all four basebands (without excluding any spectral line regions) using
CASA (version 4.7.2-REL r39762) \texttt{tclean}, and found that the central
regions surrounding Sgr B2 M were severely affected by artifacts that could not
be cleaned out.  We
therefore ran 3 iterations of phase-only self-calibration and two iterations of
amplitude + phase self-calibration, the latter using multi-scale
multi-frequency synthesis with two Taylor terms \citep{Rau2011a}, to yield a substantially
improved image (see Appendix \ref{sec:selfcal}).  The total dynamic range,
measured as the peak brightness in
Sgr B2 M to the RMS noise in a signal-free region of the combined 7m+12m image,
is 18000 (average RMS noise $0.09$ mJy/beam, 0.05 K), while the dynamic range within one
primary beam ($0.5$\arcmin) of Sgr B2 M is only 5300 (average RMS noise $0.3$
mJy/beam, 0.16 K).  Because of the dynamic range limitations and an empirical
determination that clean did not converge if allowed to go too deep, we cleaned
to a threshold of 0.1 mJy/beam over all pixels with $S_\nu > 2.5$ mJy / beam
as determined from a previous iteration of \texttt{tclean}.
The final image used for most of the analysis in this paper was imaged with 
Briggs robust parameter 0.5, achieving a beam size $0.54\arcsec\times0.46\arcsec$.
Using the same visibility data, we also produce an image with robust parameter
-1, beam size $0.37\arcsec\times0.32\arcsec$, and average RMS 0.24 mJy/beam or
0.27 K, and another tapered to exclude the long baselines imaged with robust
parameter 2 that achieved a beam size $2.35\arcsec\times1.99\arcsec$ with
average RMS 0.78 mJy/beam or 0.022 K.  All three images are distributed with
the paper (\url{https://doi.org/10.11570/17.0007}).

\Figure{f4}
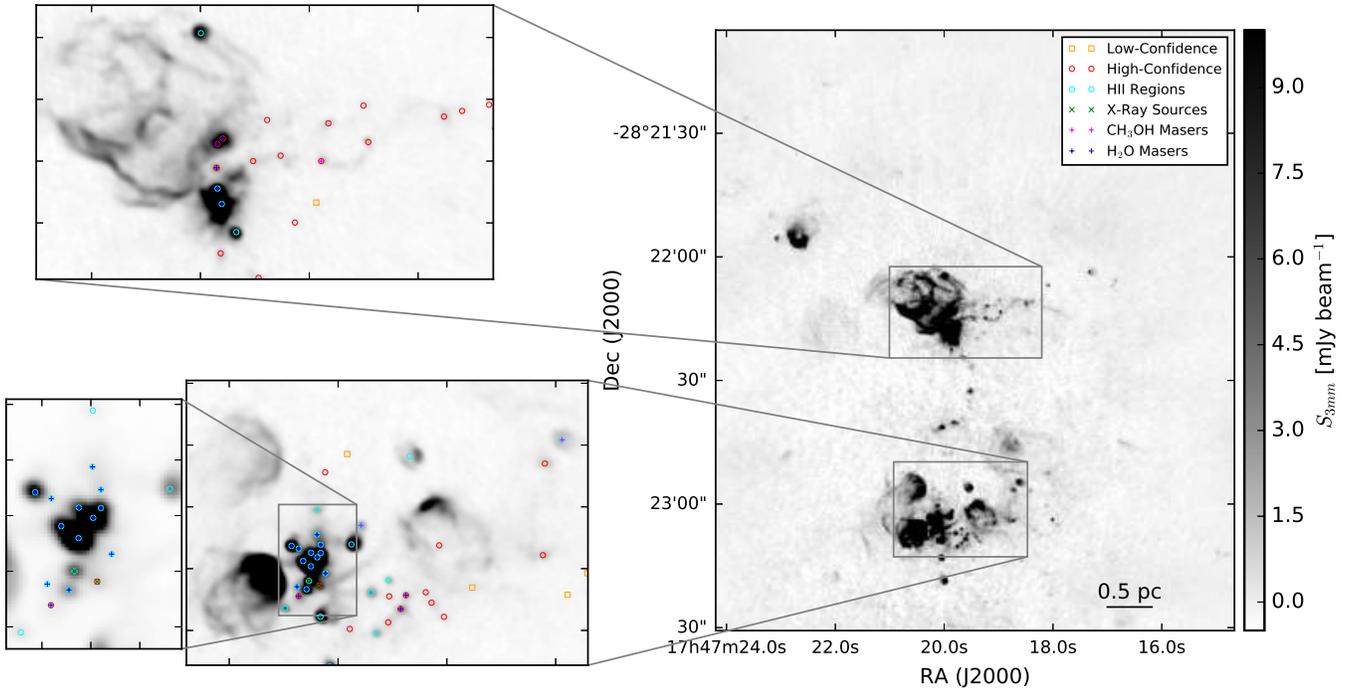
{A close-in look at the Sgr B2 M and N region.  Multiple insets show identified
sources in some of the richer sub-regions.  The points are colored as in Figure
\ref{fig:continuumMandN}.  The background image is the ALMA 3 mm continuum.
See also Figure \ref{fig:MandNzoomsVLA}.}
{fig:MandNzooms}{1}{\textwidth}

We also produced full spectral data cubes.  These were lightly
cleaned with a maximum of 2000 iterations of cleaning to a threshold of 100
mJy/beam.  The noise is typically $\approx9$ mJy \perbeam (6 K) per 0.8 \kms
channel in the robust 0.5 cubes.
No self-calibration was applied, both because the dynamic range
limitations were less significant and because the image cubes are
computationally expensive to process.
Before continuum subtraction, dynamic range related artifacts similar to those
in the continuum images were present, but these structures are nearly identical
across frequencies and were therefore removable in the image domain.  We use
median-subtracted cubes (i.e., spectral cubes with the median along each spectrum
treated as continuum and subtracted) for our analysis of the lines, noting that
the only location in which an error $>5\%$ on the median-estimated continuum is
expected
is the Sgr B2 North core \citep[][]{Sanchez-Monge2017b,Sanchez-Monge2017a}.
While many lines were included in the spectral setup\footnote{Other lines
targeted include CH$_3$CN 5-4, HCN 1-0, HNC 1-0, HCO$^+$ 1-0, H41$\alpha$, and
H$_2$CS $3_{0,3}-2_{0,2}$.} only \cyanoacetylene J=10-9 is discussed here; of
the included lines, it is the brightest and most widely detected in emission.
This line has a critical density $n_{cr}\equiv A_{ij}/C_{ij} \approx5\ee{5}$
\percc \citep{Green1978b}, so it would traditionally be considered a
high-density gas tracer.

\FigureTwo
{f5}
{f6}
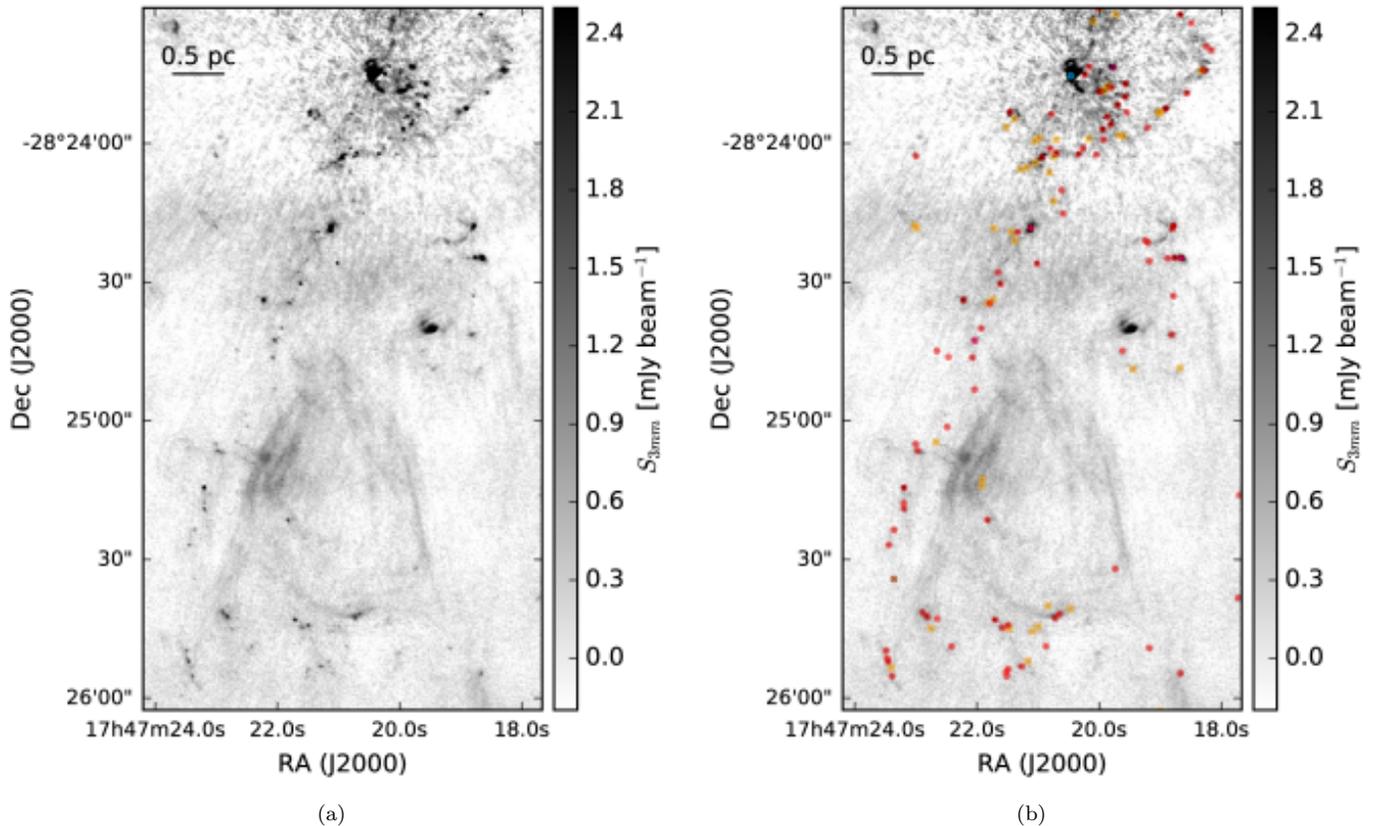
{Images of the ALMA 3 mm continuum in the Sgr B2 Deep South (DS) region.  
The right figure additionally
includes markers at the position of each identified continuum pointlike
source: red dots are `conservative', high-confidence sources,
orange squares are `optimistic', low-confidence sources,
cyan are \hii regions, magenta +'s are \methanol masers, blue +'s are \water
masers, and green X's are X-ray sources.
The \hii region Sgr B2 S is the bright source at the top of the image;
imaging artifacts can be seen surrounding it.  The largest angular
scales are noisier than the small scales; the $\sim20\arcsec$-wide east-west
ridge at around -28:24:30 is likely to be an imaging artifact.  By contrast,
the diffuse components in the southern half of the image are likely to be real.
The crowded parts of the images are shown with inset zoom-in panels
in Figure \ref{fig:deepsouthzooms}.
}
{fig:continuumDS}{1}{0.5\textwidth}

The processed data are available from \url{https://doi.org/10.11570/17.0007} in
the form of four $\sim225$ GB data cubes for the full data sets, three
continuum images at different resolutions, and two cubes of \cyanoacetylene at
different resolutions. 

\subsection{Other data - Column Density Maps}
\label{sec:colmaps}
We use archival data to create column density maps at a coarser
resolution than the ALMA data, since the ALMA data are not sensitive
enough to make direct column density measurements and because they
may be contaminated by other (non-dust) emission mechanisms.   We use Herschel
Hi-Gal data \citep{Molinari2010a} to perform SED fits to each pixel
\citep[][and in prep]{Battersby2011a}.  These fits were performed at 25\arcsec
resolution, using the 70, 160, 250, and 350 \um data and excluding the 500 \um
channel.  The estimated fit uncertainty in the column density is $25\%$, with
an upper limit on the systematic uncertainty of a factor of two (Battersby et
al, in prep).  To obtain column density maps with greater resolution, we
combine the Herschel data with SHARC 350 \um and SCUBA 450 \um images.

The CSO SHARC data were reported in \citet{Bally2010a} and have a nominal
resolution of 9\arcsec at 350 \um, however, at this resolution, the SHARC data
display a much higher surface brightness than the Herschel data on the same
angular scale.  An assumed resolution of 11.5\arcsec gives a better surface
brightness match and is consistent with the measured size of Sgr B2 N in the
image.  This calibration difference is likely to have been produced by a
combination of blurring by pointing errors, surface imperfections, and the
gridding process, all of which increase the effective beam size, and flux
calibration errors.  In any case, the Herschel data provide the most
trustworthy absolute calibration scale, since they were taken from space and
calibrated to an absolute scale using Planck data
\citep{Bendo2013a,Bertincourt2016a}, so we assume the Herschel calibration is
correct when combining the data.

The JCMT SCUBA 450 \um data were reported in \citet{Pierce-Price2000a} and
\citet{di-Francesco2008a} with a resolution of 8\arcsec.  We found that the
SCUBA data had a flux scale significantly discrepant from the Herschel-SPIRE
500 \um data on 30-90\arcsec scales, even accounting for the central wavelength
difference.  We had to scale the SCUBA data up by a factor $\approx3$ to make
the data agree with the Herschel-SPIRE images on these scales.  While such a
large flux calibration error seems implausible, it can occur if the beam size
of the ground-based data is larger than expected.  To assess this possibility,
we fit 2D Gaussians to several sources in the SCUBA CMZ maps, measuring a FWHM
toward Sgr B2 N of approximately 14\arcsec (and toward several other sources,
$>10.5\arcsec$), which means the observed beam area is about three times larger
than theoretically expected.
Between the larger beam area, flux calibration errors \citep[quoted at
20\percent in][]{Pierce-Price2000a}, and the dust emissivity correction
(35-50\percent for dust index $\beta=1-2$, where $\beta=\alpha-2$), this large
$3\times$ flux scaling factor is plausible.  The large secondary error beam
\citep[17.3\arcsec][]{di-Francesco2008a} of the 450 \um SCUBA data may also
contribute to this effect.  As with the SHARC data above, we trust the
space-based calibration over the ground-based.

We combined the Herschel data with the SHARC and SCUBA data to create
higher-resolution maps at 350 \um (Herschel-SPIRE+SHARC) and 450 \um
(Herschel-SPIRE+SCUBA).  The data combination process is discussed in detail in
Appendix \ref{sec:singledishcomb}, but in brief, we used a `feathering'
technique \citep[e.g.,][as implemented in
\texttt{uvcombine}\footnote{https://github.com/radio-astro-tools/uvcombine}]{Stanimirovic2002a,Cotton2017a}
to combine the images in the Fourier domain.

Using these higher-resolution maps, we created several column density maps
using different assumptions about the dust temperature.  For simplicity, we
produced maps assuming arbitrary constant temperatures equal to the minimum and
maximum expected dust temperatures (20 and 50 K). We produced additional maps
using the temperature measured with Herschel SED fits interpolated onto the
higher-resolution SCUBA and SHARC grids.  Because of the interpolation and
fixed temperature assumptions, the column maps are not very accurate and should
not be used for systematic statistical analysis of the column density
distribution (i.e., PDF shape analysis) without careful attention to the large
implied uncertainties.  However, these higher-resolution data are used in this
paper to provide the best estimates of the local column density around our
sample of compact millimeter continuum sources.

One important uncertainty in these column density maps is possible foreground
or background contamination.  Sgr B2 is \dsgrb away from us in the direction of
our Galaxy's center, meaning there is a potentially enormous amount of material
unassociated with the Sgr B2 cloud along the line of sight.  This material may
have column densities as low as 5\ee{21} \persc or as high as 5\ee{22} \persc,
as measured from relatively blank regions in the Herschel column density map
\citep[][and in prep]{Battersby2011a}.  The former value corresponds to the
background at high latitudes, $b\sim0.5$, while the latter  is approximately
the lowest seen within our field of view. 

\section{Analysis of the continuum sources}
\label{sec:analysis}
In this section, we identify continuum sources (\S \ref{sec:contsources}),
match them with other catalogs (\S \ref{sec:crossmatch}), and discuss
their nature (\S \ref{sec:classification}).

\subsection{Continuum Source Identification}
\label{sec:contsources}
We selected compact continuum  sources by eye,
scanning across images with different weighting schemes (different robust
parameters).  An automated selection is not viable across the majority of the
observed field for several reasons:
\begin{enumerate}
    \item There are many extended \hii regions that dominate the overall map
        emission.  These are clumpy and have local peaks that would dominate
        the identified source population using most source-finding algorithms.
    \item There are substantial imaging artifacts produced by the extremely
        bright emission sources in Sgr B2 M ($S_{3 \textrm{mm,max}} \approx 1.6$ Jy) and
        Sgr B2 N ($S_{3 \textrm{mm,max}} \approx 0.3$ Jy) that make automated source
        identification particularly challenging in the most source-dense
        regions.  These are `sidelobes' from the bright sources that cannot be
        entirely removed.
    \item Resolved-out emission has left multi-scale artifacts throughout the
        images.  While these can be filtered out to a limited degree by
        excluding large angular scales (short baselines), there remain
        small-scale ripples, and the noise increases when baselines are
        excluded.
\end{enumerate}
All of these features are evident in Figures \ref{fig:continuumMandN} and
\ref{fig:continuumDS}. 

Because the noise varies significantly across the map (it is higher near Sgr B2
M), and because there is extended emission, a uniform selection criterion is
not possible.  We therefore include two levels of source identification,
`high-confidence' sources, which are peaks clearly above the noise in regions
of low-background, and `low-confidence' sources that are somewhat lower
signal-to-noise  and are often in regions with higher background in which the
noise estimate may be inaccurate.  The difference between the high- and low-
confidence sources is subjective, since it is based in part on a by-eye
assessment of how much the local noise is affected by resolved-out structure.
Part of the by-eye assessment involved blinking between the three images with
different resolution described in \S \ref{sec:observations}; if a structure
looked point-like in the highest-resolution image, but turned out to be part of
a more extended structure in the lowest-resolution (and highest-sensitivity)
image, we marked it as `low-confidence'.

Outside of the dense clusters, every peak that is higher than five times the
lowest measured RMS noise value was visually inspected.  Peaks that were part
of extended structures but not significantly different from them (e.g., a
5-$\sigma$ peak sitting on a 4-$\sigma$ extended structure) were excluded.  We
excluded sources with radial extents $r>1\arcsec$ ($r>0.04$ pc), i.e., extended
\hii regions (all such sources have corresponding centimeter-wavelength
detections indicating that they are \hii regions).

We measure the local noise for each source by computing the median absolute
deviation in an annulus 0.5 to 1.5\arcsec around the source center; these noise
measurements are reported in Table~\ref{tab:photometry}.

Our selection criteria result in a reliable but potentially incomplete catalog;
because we did not employ an automated source identification algorithm, we
cannot readily quantify our completeness.  The regions most likely
to be incomplete near our noise threshold are Sgr B2 M and N.  In these
regions, dynamic range limitations increase the background noise and make
fainter sources difficult to detect, as described in Section
\ref{sec:observations}.  Additionally, they both contain extended structures,
including \hii regions and dust filaments, which likely obscure compact
sources.

For a subset of the sources, primarily the brightest, we measured the spectral
index $\alpha$ based on CASA \texttt{tclean}'s  2-term Taylor expansion model
of the data (parameters \texttt{deconvolver=`mtmfs'} and \texttt{nterms=2}).
This measurement is over a narrow frequency range ($\approx90-100$ GHz).
\texttt{tclean} produces $\alpha$ and $\sigma(\alpha)$ (error on $\alpha$)
maps, and we used the $\alpha$ value at the position of peak intensity for each
source.  We include in the analysis only those sources with $|\alpha| > 5
\sigma(\alpha)$ or $\sigma(\alpha) < 0.1$; the latter include sources with
$\alpha\sim0$ measured at relatively high precision.  We exclude the lower-precision
measurements of $\alpha$ because they are not useful for identifying the emission
mechanism.  Of the \ncores detected sources, \nalphas met these
criteria. Several of the brightest sources did not have significant
measurements of $\alpha$ because they are in the immediate neighborhood of Sgr
B2 M or N and therefore have significantly higher background and noise,
preventing a clear measurement.

To check the calibration of the spectral index measurement, we imaged one of
our calibrators, J1752-2956, and obtained a spectral index
$\alpha=-0.62\pm0.14$, consistent with the expected $\alpha\approx-0.7$ for an
optically thin synchrotron source \citep[e.g.,][]{Condon2007a}.  We also note
that the \emph{relative} spectral index measurements in our catalog should be
accurate, since all sources come from the same map with identical calibration.

\Figure{f7}
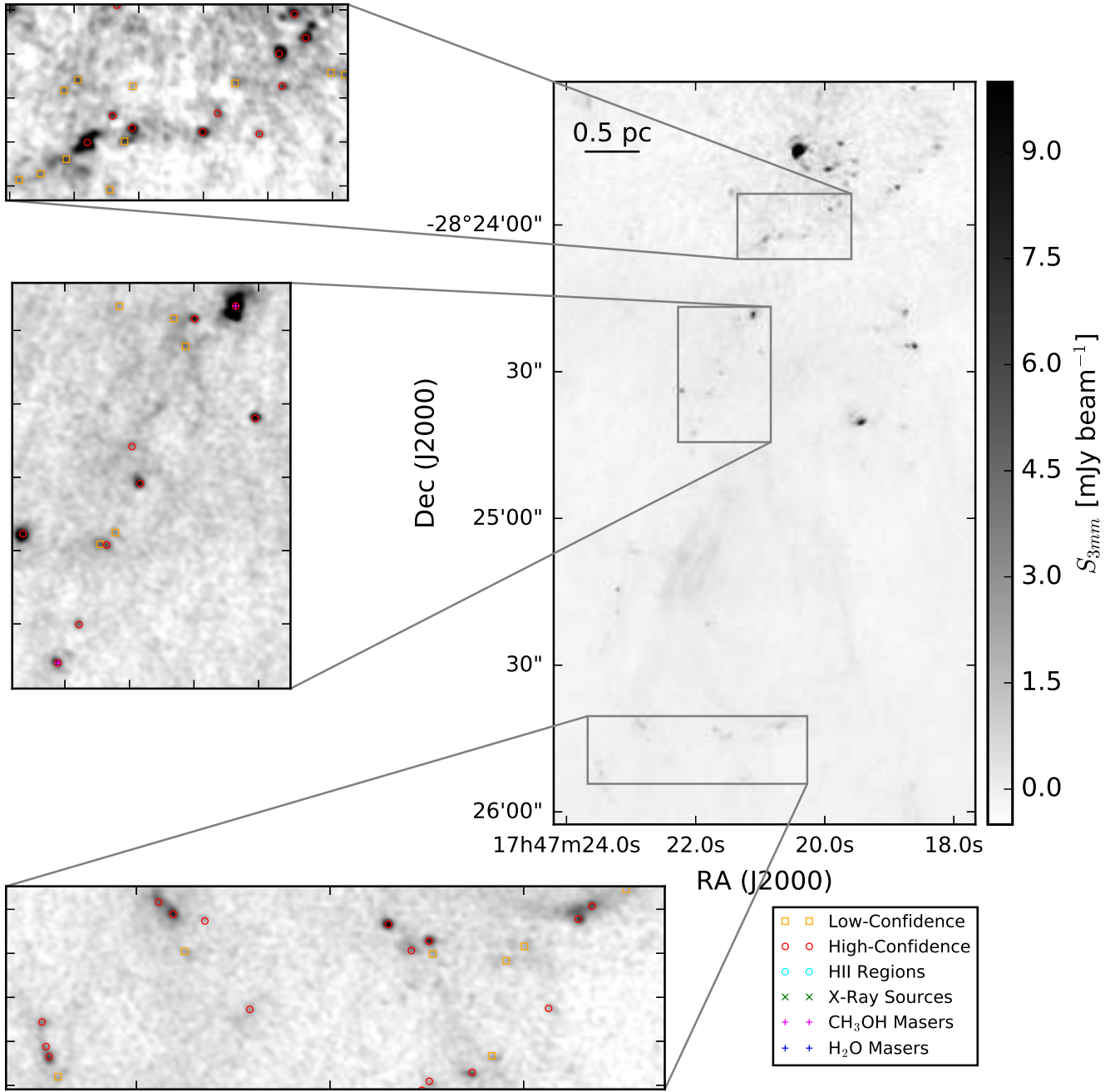
{A close-in look at the Sgr B2 DS region.  Multiple insets show identified sources
in some of the richer sub-regions.  The points are colored as in Figure
\ref{fig:continuumMandN}.  The background image is the ALMA 3 mm continuum.}
{fig:deepsouthzooms}{1}{\textwidth}

We detected \ncores compact continuum sources, and they are listed
in Table~\ref{tab:photometry}.  Their flux distribution is
shown in Figure \ref{fig:fluxhist}.  The distribution of their measured
spectral indices $\alpha$ is shown in Figure \ref{fig:alphahist}.
Generally, spectral indices $\alpha<0$ indicate nonthermal (e.g., synchrotron)
emission, $-0.1<\alpha<2$ may correspond to free-free sources of various
optical depths, $\alpha=2$ for any optically thick thermal source,
and $\alpha>2$ usually indicates optically thin dust emission.  These indices
will be discussed further in Section \ref{sec:classification}.

\subsection{Source Classification based on Catalog Cross-Matching}
\label{sec:crossmatch}
We cross-matched our source catalog with catalogs of \ammonia sources, \hii
regions, X-ray sources, Spitzer sources, and methanol and water masers.

\paragraph{\hii regions}
We classified sources as \hii regions if there is a corresponding 0.7 or 1.3 cm
source from one of the previous VLA surveys
\citep{Gaume1995a,Mehringer1995b,De-Pree1996a,De-Pree2015a} within one ALMA
beam (0.5\arcsec).  \nhii of our sources are classified as \hii regions; these
all have $S_{3 \textrm{mm}} > 9$ mJy.  The
majority of these are unresolved, but we have included \hii regions with radii
up to $r\leq1\arcsec$ in our catalog.  Optically thick \hii regions (like any
blackbody) have a spectral index $\alpha=2$.  Optically thin \hii regions have
a nearly flat spectral index, $\alpha=-0.1$ \citep{Condon2007a}.   The observed
sources with \hii region counterparts have spectral indices consistent with the
theoretical expectation for optically thin \hii regions in Figure
\ref{fig:alphahist}.  The existing VLA data do not cover the entire area of our
observations, so we only have a lower limit on the number of \hii regions in
our sample; the sources in Sgr B2 DS have not yet been observed in the radio at
high resolution.  Sources matched with \hii regions evidently contain high-mass
(most likely $M\gtrsim20$ \msun, see Section \ref{sec:theyarehiiregions} below)
young stars.

\paragraph{\ammonia sources}
\citet{Martin-Pintado1999a} observed part of Sgr B2 DS and M in \ammonia with
the VLA.  They identified three ``hot cores'' based on \ammonia (4,4) detections.
Only their first source HC1 has an associated 3 mm continuum source, suggesting
that HC2 and HC3 are not genuine hot cores but are some other variant of
locally heated (perhaps shock-heated) gas.  However, the
association between HC1 and our source 43 suggests that it is a YSO with a
massive envelope.  Of the 6 \ammonia (3,3) maser sources identified
by \citet{Martin-Pintado1999a}, three are in regions with high 3 mm source
density but lack a clear one-to-one source association, one is coincident with
an extended \hii region not in our catalog, and two have no obvious
associations.  The \ammonia (3,3) masers therefore do not appear to be
unambiguous tracers of star formation in this environment, consistent with the
conclusions of \citet{Mills2015a}.

\paragraph{6.67 GHz \methanol masers}
Class II methanol masers are exclusively associated with sites of high-mass
star formation.  The \citet{Caswell2010a} Methanol Multibeam (MMB) Survey
identified 11 sources in our observed field of view (their survey covers our
entire observed area), of which \nmasermatch have a clear match to within
1\arcsec of a source in our catalog (the MMB catalog sources have a positional
accuracy of $\approx0.4\arcsec$, but masers may have an extent up to 1\arcsec).
These sources are clearly identified as high-mass YSOs.
The single maser that does not have an associated millimeter source is 5\arcsec
west of Sgr B2 S and resides near some very faint and diffuse 3 mm
emission; it is unclear why the 3 mm is so weak here, but it hints that there
are MYSOs with 3 mm emission below our
detection limit.

\paragraph{\water masers}
Water masers are generally associated with young, accreting stars.  We
matched our catalog with the \citet{McGrath2004a} water maser catalog,
finding that 23 of our sources have a water maser within 1\arcsec.
These sources are likely to contain YSOs, but not necessarily
MYSOs based on their \water maser detections alone.  There are 14 masers
from their catalog that do not have associated sources in our catalog, though
not all of these maser spots are spatially distinct.  Most of these unassociated
masers are seen outside of Sgr B2 N and Sgr B2 S and may be associated with
outflows.  This catalog covers about 10\% of our mapped area with their
single VLA K-band pointing; their map excludes the many sources in Sgr B2 DS.

\paragraph{X-ray sources}
Some young stars exhibit X-ray emission, including some MYSOs
\citep[e.g.,][]{Townsley2014a}, so we searched for X-ray emission from our
sources.  \nxraymatch of the sources have X-ray counterparts in the
\citet{Muno2009a} Chandra point source catalog within 1\arcsec.  The
\citet{Muno2009a} catalog covers our entire observed area.  The X-ray
associated sources most likely contain YSOs.  There are 102 X-ray sources in
the field of view that do not have associated 3 mm sources.

\paragraph{Spitzer mid-infrared sources}
We searched the \citet{Yusef-Zadeh2009a} catalogs of 4.5 \um excess sources and
YSO candidates and found only one source association, though there are 5 and
14, respectively, of these sources in our field of view.  Two of the 4.5 \um
excess sources and one of the YSO candidates are associated with extended \hii
regions (which we do not catalog); the single association is of a 4.5 \um source with the central region
of Sgr B2 M. By-eye comparison of the Spitzer maps and the ALMA images suggests
that the lack of associations is at least in part because of the high
extinction in the regions containing the 3 mm cores; there are overall fewer
Spitzer sources in these parts of the maps.

\paragraph{44 GHz \methanol masers}
Finally, we searched the \citet{Mehringer1997a} sample of 44 GHz Class I
\methanol maser sources for associations, finding no matches with any of our
sources out of the 18 nonthermal \methanol emission sources they reported.
This methanol maser line apparently does not trace star formation.
Their maps include two VLA Q-band images pointed at Sgr B2 M and N; these
maps cover only a very small fraction ($\sim5\%$) of our mapped area.

\subsection{Nature of the Continuum Sources}
\label{sec:classification}
The majority of the detected sources are observed only as 3 mm continuum
sources, with no spectral line information or detection at other wavelengths.
In this section, we employ a variety of arguments to classify the sample of new
sources.    Plausible emission mechanisms include free-free and thermal dust
emission, so in this section we explore whether the sources could be different
classes of dust or free-free sources.  We examine whether they are
prestellar cores (\S \ref{sec:alt0}), externally ionized globules (\S
\ref{sec:alt1}), \hii regions from an extended population of OB stars (\S
\ref{sec:alt2}), or \hii regions around young massive stars (\S
\ref{sec:theyarehiiregions}).  After determining that the above alternatives do
not readily explain the whole sample, we conclude that the sources are
primarily dense gas and dust cores with internal heating sources (\S
\ref{sec:theyareprotostars}).

\paragraph{A lack of line emission}
We visually inspected the spectra extracted from the full line cubes, and no
lines are detected peaking toward most of the sources (most sources
have emission in some lines, such as \cyanoacetylene 10-9, but this emission is
clearly extended and not associated with the compact source).  Given the
relatively poor line sensitivity (RMS $\approx 6$ K), the dearth of detections
is not very surprising.  We therefore cannot use spectral lines to classify
most sources.

\FigureOneCol{f8}
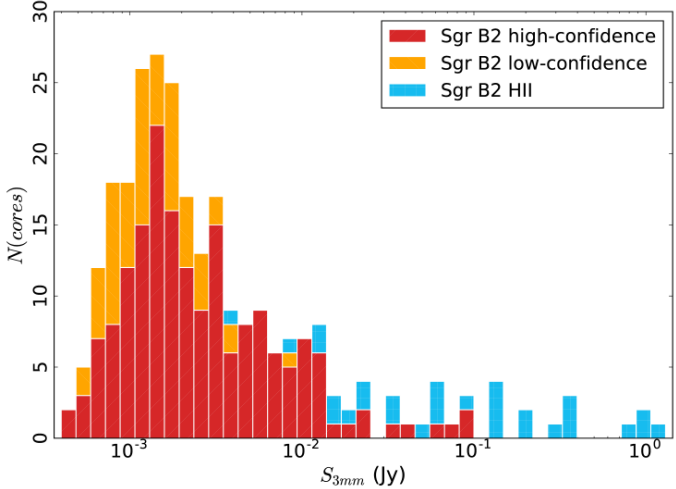
{A histogram of the flux density (the peak intensity converted to flux density
assuming the source is unresolved) of the observed sources. 
The histograms are stacked such that there are a total of 27 sources in the
highest bin.
}
{fig:fluxhist}{1}{0.5\textwidth}

\subsubsection{Alternative 1: The sources are `prestellar' cores}
\label{sec:alt0}
The simplest assumption is that all sources we have detected that were not
detected at longer wavelengths are pure dust emission sources at a constant
temperature, i.e., they are starless cores.

At \dsgrb, a 1 mJy source corresponds to an optically thin gas mass\footnote{We
assume a gas-to-dust ratio of 100 throughout this work.} of
$M(40\mathrm{K})=18$ \msun or $M(20\mathrm{K})=38$ \msun assuming a dust
opacity index $\beta=1.75$ (spectral index $\alpha=3.75$ if measured on the
Rayleigh-Jeans tail of the spectral energy distribution) to extrapolate the
\citet[][MRN with thin ice mantles anchored at 1mm]{Ossenkopf1994a} opacity to
$\kappa_{3.1 \mathrm{mm}}=0.0018$ cm$^2$ g$^{-1}$ (per gram of gas).  Our
dust-only (i.e., excluding free-free emission) 5-$\sigma$ sensitivity limit at
20 K therefore ranges from $M>19$ \msun (0.5 mJy) to $M>94$ \msun (2.5 mJy)
across the map.  If we were to assume that these are all cold, dusty sources,
as is typically (and reasonably) assumed for local clouds, they would be
extremely massive and dense, with the lowest measurable density being
$n(20\mathrm{K}) > 1\ee{8}$ \percc (corresponding to 19 \msun in an
$r=0.2$\arcsec$=1700$ AU radius sphere, i.e., a sphere with radius equal to the
beam $1-\sigma$ size).

Such extreme objects are technically possible, but we argue the majority are
unlikely to fall into this class.  We have detected $>100$ of these sources,
but only a handful of comparable-mass starless cores have ever been claimed
before \citep[e.g.,][]{Kong2017a}, and few of those reported are so compact
\citep[e.g.,][]{Cyganowski2014a}.   Theoretical models of high-mass prestellar
cores \citep{McKee2003a} suggest they are much larger and less concentrated
than the sources we observe.

At the high implied densities (${n(20 \mathrm{K})>10^8~\percc}$), it is
unlikely that the cores are unbound; these sources have ${v_{esc} > 2~\kms
(M/10~ \msun)^{1/2}}$ from ${r=0.5\arcsec=4200\mathrm{~AU}}$.  The high density
required for our sources results in a short free-fall timescale, ${t_{ff}<3000
(n/10^8 ~\percc)^{-1/2}\mathrm{~yr}}$.  Assuming such cores do exist, the
timescale for them to form a central YSO (a central heating source) is short.
While there are few constraints on the accreting lifetime of high-mass YSOs,
that timescale is almost certainly 1-2 orders of magnitude longer.  For a given
population of cores, we would expect only of order 1-10\% of them to be
starless at any given time.  We will revisit the characteristics of centrally
heated dust sources in Section \ref{sec:theyareprotostars} below.

\subsubsection{Alternative 2: The sources are externally ionized gas blobs}
\label{sec:alt1}
One possibility is that these sources are not dust-dominated, nor pre- or
protostellar, but are instead externally ionized, mostly neutral gas clumps
embedded within diffuse \hii regions.  They would then be analogous to the
heads of cometary clouds, externally ionized globules
\citep[``EGGs"; ][]{Sahai2012a}, or proplyds (externally ionized protoplanetary
disks), and their observed emission would give little clue to their nature because
the light source is extrinsic.

The majority of the detected sources have size $<2000$ AU, i.e., they are
unresolved\footnote{We consider a source unresolved if its radius is smaller
than the Gaussian width of our beam, $0.2\arcsec\approx2000$ AU, rather than
the FWHM of $0.5\arcsec\approx4000$ AU, since a source with the latter width
would be measurably extended when convolved with the beam.}.  By contrast, the
free-floating EGGs (`frEGGs') so far observed have sizes 10,000-20,000 AU
\citep{Sahai2012a,Sahai2012b}, so they would be resolved in our observations.
Toward the brightest frEGG in Cygnus X, \citet{Sahai2012b} measured a peak
intensity $S_{8.5 GHz} \approx 1.5$ mJy/beam in a $\approx3\arcsec$ beam.
Cygnus X is $6\times$ closer that the Galactic center, so their beam size is
the same physical scale as ours.  If the free-free emission is thin
($\alpha=-0.1$), the brightness in our data would be $S_{95 GHz} =
(95/8.5)^{-0.1} S_{8.5 GHz} = 0.79 S_{8.5 GHz} \approx 1.2$ mJy/beam.  These
frEGGs would be detectable in our data.  Comparison to radio observations at a
similar resolution will be needed to rule out the externally ionized globule
hypothesis for the resolved regions within our sample.  However,  the
unresolved sources in our sample are unlikely to be frEGGs, since they are
too small.

\FigureOneCol{f9}
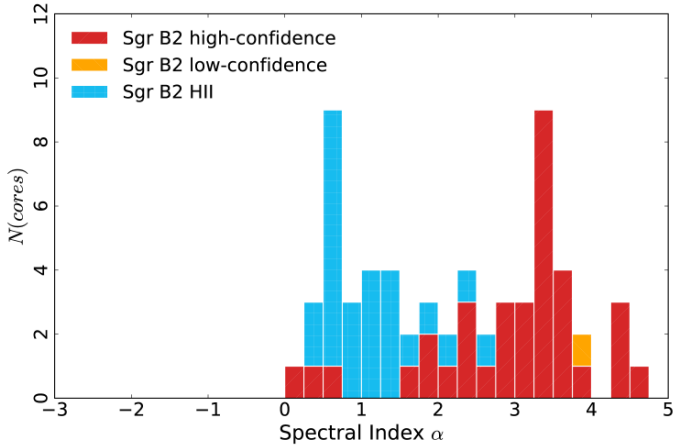
{A histogram of the spectral index $\alpha$ for those sources with a statistically
significant measurement.  The \hii regions cluster around $\alpha=0$, as expected
for optically thin free-free emission, while the unclassified sources cluster
around $\alpha=3.5$, which is  consistent with dust emission.
}
{fig:alphahist}{1}{0.5\textwidth}

If the detected sources were either EGGs or cometary clouds, we would expect
them to be located within diffuse \hii regions, since that is where all other
sources of this type are seen, and since an external ionizing agent is needed
to illuminate them.  Many of the sources are near, but not embedded in, \hii
regions, as seen in Figure \ref{fig:coreson20cmandhc3n}a, which shows 20 cm
continuum emission that most likely traces ionized gas.  The sources are
nearly all associated with a ridge of molecular (\cyanoacetylene) emission
(Figure
\ref{fig:coreson20cmandhc3n}b).  If they are deeply embedded within the
molecular material, they cannot be externally ionized.  

The ionized gas emission (20 cm, Figure \ref{fig:coreson20cmandhc3n}a) and
molecular gas emission (\cyanoacetylene, Figure \ref{fig:coreson20cmandhc3n}b)
are anticorrelated.  The \cyanoacetylene emission wraps around the 20 cm
emission, and has a significant extent beyond the edge of the 20 cm emission.
If the \cyanoacetylene were tracing a photon-dominated region, we would expect
the \cyanoacetylene emission to peak along the edge of the \hii region.
Since it does not, we conclude that the \cyanoacetylene emission is tracing a
`quiescent' molecular cloud, i.e., one that is not significantly heated by
the adjacent \hii region.  Most of the 3 mm sources are aligned with bright
\cyanoacetylene emission, implying that they are embedded within it.
If they are indeed embedded in an extended molecular cloud, that cloud
should shield them from ionizing radiation.  The sources are therefore
mostly not externally ionized.

A final point against the externally ionized hypothesis is the observed
spectral indices shown in Figure \ref{fig:alphahist}.  We measured spectral
indices for \nalphas sources, of which \ngttwo have $\alpha>2$.  These \ngttwo
sources are inconsistent with free-free emission and are at least reasonably
consistent with dust emission.

\FigureTwo
{f10}
{f11}
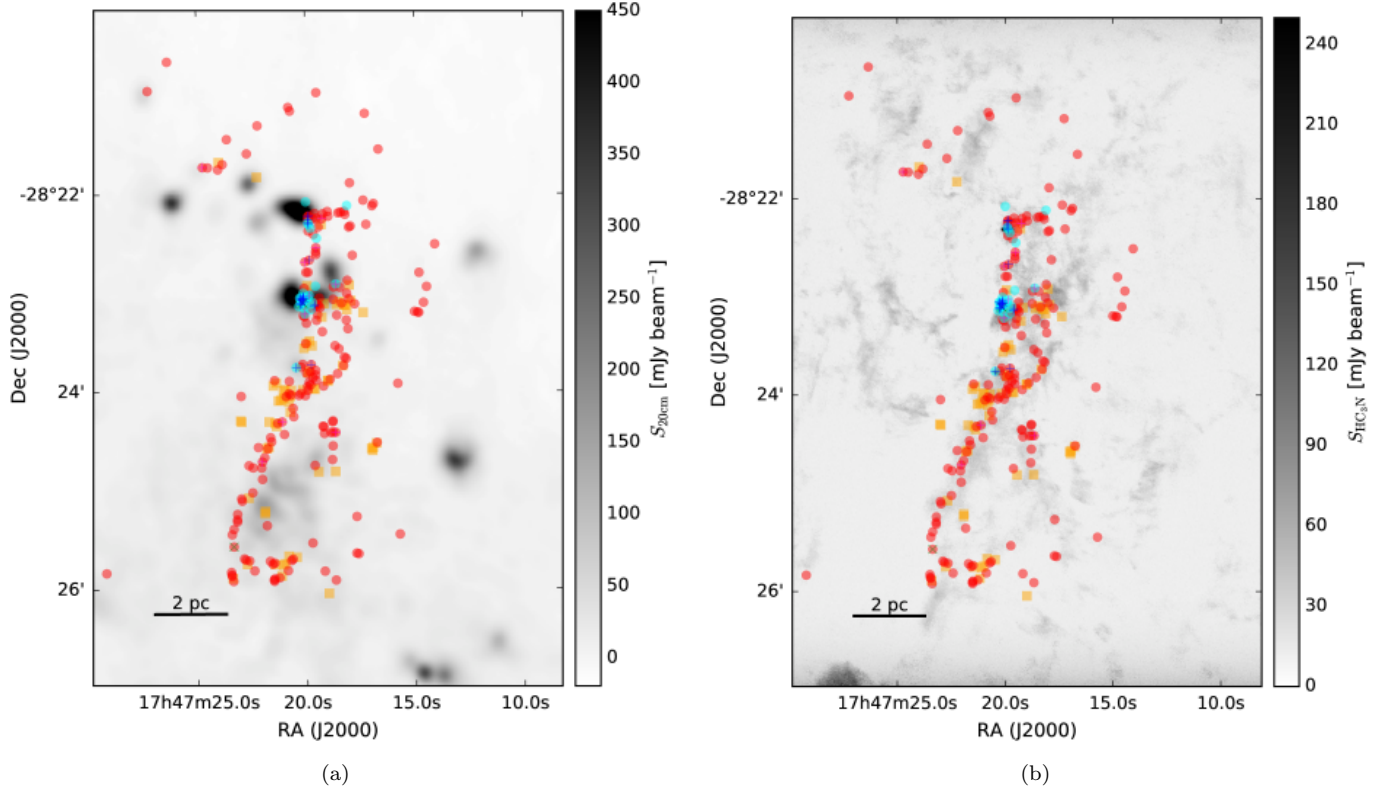
{(left) The location of the detected continuum sources (red points) overlaid on a 20
cm continuum VLA map highlighting the diffuse free-free (or possibly
synchrotron) emission in the region \citep{Yusef-Zadeh2004a}.
(right) Continuum sources overlaid on a map
of the \cyanoacetylene J=10-9 peak intensity over the range [-200, 200] \kms.
  In both figures, red
dots are `conservative',
high-confidence sources, orange squares are `optimistic', low-confidence sources,
cyan are \hii regions, magenta +'s are \methanol masers, blue +'s are \water
masers, and green X's are X-ray sources.
}
{fig:coreson20cmandhc3n}{1}{0.5\textwidth}

\subsubsection{Alternative 3: The sources are \hii regions produced by
interloper ionizing stars}
\label{sec:alt2}
If there is a large population of older (age 1-30 Myr) massive stars, they
could ignite compact \hii regions when they fly through molecular material.  In
other words, each OB star that encounters dense enough gas would create a
compact \hii region that would not have time to expand due to the star's rapid
motion.  Such sources would be bow-shaped when viewed at higher resolution.
See \S \ref{sec:theyarehiiregions} for calculations of stationary \hii region
properties.

The main problem with this scenario is the spatial distribution of the observed
sources.  While most of the continuum sources are associated with dense gas and
dust ridges, not all of the high-column-density molecular gas regions have such
sources in them (see Figure \ref{fig:coreson20cmandhc3n}b, where there is some
molecular material that does not have associated millimeter sources,
especially to the east and west of the main ridge).  If there is a
free-floating population of OB stars responsible for the 3 mm compact source
population, and if we assume the spatial distribution of the stars is uniform,
the distribution of the resulting \hii regions should match that of the gas.
Also, there is no such population of sources seen \emph{outside} of the dense
gas in the infrared,
which again we should expect if there is a uniformly distributed massive stellar
population.  Finally, the spectral indices discussed above (Figure
\ref{fig:alphahist}) suggest the previously-unidentified sources are dust
emission sources, not free-free sources.

\subsubsection{Alternative 4: The sources are \hii regions produced by
recently-formed OB stars}
\label{sec:theyarehiiregions}

We know from previous observations
\citep[e.g.,][]{Mehringer1995b,De-Pree1996a,De-Pree2015a} that there is a
substantial population of \hii regions in the Sgr B2 clusters.  The \nhii
sources associated with these previously-identified \hii regions are among the
brightest in our catalog.  We address here whether the remaining  sources,
which are mostly fainter, could also be \hii regions.

\begin{sloppypar}
To calculate the expected 3 mm flux density from an \hii region with a central
source emitting Lyman continuum luminosity $Q_{lyc}$, we rearrange
\citet{Condon2007a} equations 4.60 and 4.61.  We get an equation for the
expected brightness temperature as a function of electron temperature $T_e$,
emission measure $EM$, and frequency $\nu$:
\begin{subequations}
    \label{eqn:hiiregion}
    \begin{align}
T_B &= T_e  \left[1-\exp\left(-\tau\right) \right] \\
\tau &= c_* T_* \nu_* EM_* \\
\nu_* &= \left(\frac{\nu}{\mathrm{GHz}}\right)^{-2.1} \\
T_* &= \left(\frac{T_e}{10^4 \mathrm{K}}\right)^{-1.35} \\
c_* &= -3.28\times10^{-7} \\
EM &= \frac{3 Q_{lyc}}{4 \pi R^2 \alpha_B} \\
EM_* &= \frac{EM}{\mathrm{pc~cm}^{-6}} 
    \end{align}
\end{subequations}
where 
$Q_{lyc}$ is the count rate of ionizing photons in $s^{-1}$,
$\tau$ is the optical depth of the \hii region,
${\alpha_B=2\ee{-13}\mathrm{~cm}^3 \mathrm{s}^{-1}}$ 
is the case-B recombination coefficient,
and $R$ is the \hii region radius.  The emission measure $EM_*$ assumes the
\hii region is a uniform-density Str{\"o}mgren sphere.  The constant $c_*$ was
computed by \citet{Mezger1967a} as an approximation to the optical depth
prefactor in the full radiative transfer equation and is never incorrect by
more than $\approx25\%$.
To convert the above brightness temperature into a flux density, assuming a
${\mathrm{FWHM}=0.5\arcsec}$ beam at 95 GHz, ${1\mathrm{~K} = 1.85
\mathrm{~mJy~beam}^{-1}}$.
\end{sloppypar}

For an unresolved spherically symmetric \hii region (${R=4000 \mathrm{~AU}}$),
the expected flux density is ${S_{95 \mathrm{GHz}} = 5.2 \mathrm{~mJy}}$ for a
${Q_{lyc}=10^{47}~\pers}$ source (assuming ${T_e=7000\mathrm{~K}}$), and that
value scales linearly with $Q_{lyc}$ as long as the source is optically thin
(in the optically thin ${\tau\ll1}$ limit, equation \ref{eqn:hiiregion}a 
becomes approximately ${T_B=\tau T_e}$).

An extremely compact \hii region, e.g., one with $R<100$ AU and corresponding
density $n>10^6$ \percc, would be somewhat optically thick ($\tau\approx0.65$)
and therefore fainter, ${S_{95 \mathrm{GHz}}(R=100 \mathrm{AU}, Q_{lyc}=10^{47}
\pers)=3.4\mathrm{~mJy}}$.  Even the most luminous O-stars could produce \hii regions as
faint as 0.5 mJy if embedded in extremely high density gas; above
$Q_{lyc}>10^{47}$ \pers, a 25 AU \hii region would have $S_{95 GHz}\approx0.5$
mJy ($\tau=10$).

Figure \ref{fig:hiibrightness} shows the predicted brightness for various \hii
regions produced by OB stars and the density required for those \hii regions
to be the specified size.  There is a narrow range of late O/early B\footnote{We use
the tabulations of OB star properties from \citet{Vacca1996a} and
\citet{Pecaut2013a}, via their online table
\url{http://www.pas.rochester.edu/~emamajek/EEM_dwarf_UBVIJHK_colors_Teff.txt},
to determine the relation between spectral type, luminosity, and mass.} stars,
$10^{46} < Q_{lyc} < 10^{47}$ \pers, that could be embedded in compact \hii
regions of almost any size and produce the observed range of flux densities.
In order for the detected sources to be O-star-driven \hii regions, with $10^{47}
< Q_{lyc} < 10^{50}$ \pers, they must be optically thick and therefore
extremely compact and dense.  
Anything fainter, i.e., later than $\sim$B2 ($Q_{lyc}<10^{46}$ \pers), would be
incapable of producing the observed flux densities.

The 119 sources with $1.5 \mathrm{~mJy} < S_{95 GHz} < 10$ mJy that were not
previously identified as \hii regions from radio data require a finely tuned set of
parameters to be \hii regions.  Stars emitting $5\ee{46} < Q_{lyc} < 2\ee{47}$
photons per second (B1.5-B2 main sequence stars, with $M\approx8-10$ \msun) could
reside in \hii regions spanning a wide range of radii and produce flux
densities in the observed range (Figure \ref{fig:hiibrightness}a).  More
luminous stars could reside in 50-100 AU \hii regions and produce the observed
flux densities, but such small regions
are expected to be very short-lived and therefore rare.  It is unlikely that
nearly half of the stars are between 8-10 \msun, since such a local mass peak
would imply a highly abnormal IMF\footnote{Assuming all 50 sources with
$S_{3mm}>10$ mJy are massive stars with $M>10$ \msun, only 17 stars in the range
8-10 \msun are expected assuming a \citet{Kroupa2001a} IMF.}.  We therefore
assume that the newly detected sources are not predominantly \hii regions.

For completeness, we assess the emission properties of the dust surrounding
hypercompact \hii regions, since, in order to remain hypercompact, the stars
must be surrounded by very dense gas.  Figure \ref{fig:hiibrightness}b shows
that, if O-stars were confined to \hii regions small enough to produce the
median source flux density (2 mJy), the emission could be dominated by a
surrounding warm (40 K) dust core.  Such sources would be at least twice as
bright as predicted in Figure \ref{fig:hiibrightness}a.  Only the most luminous
O-stars are affected by this consideration, however, this plot also illustrates
that O-stars will almost certainly be detected in our data no matter how dense
their surroundings.

A final point against the sample being exclusively \hii regions is the observed
spectral indices.  While some are consistent with \hii regions, with $\alpha <
2$, some ($\ngttwo$) are steeper than $\alpha>2$ and are therefore inconsistent
with free-free emission.

\Figure{f12}
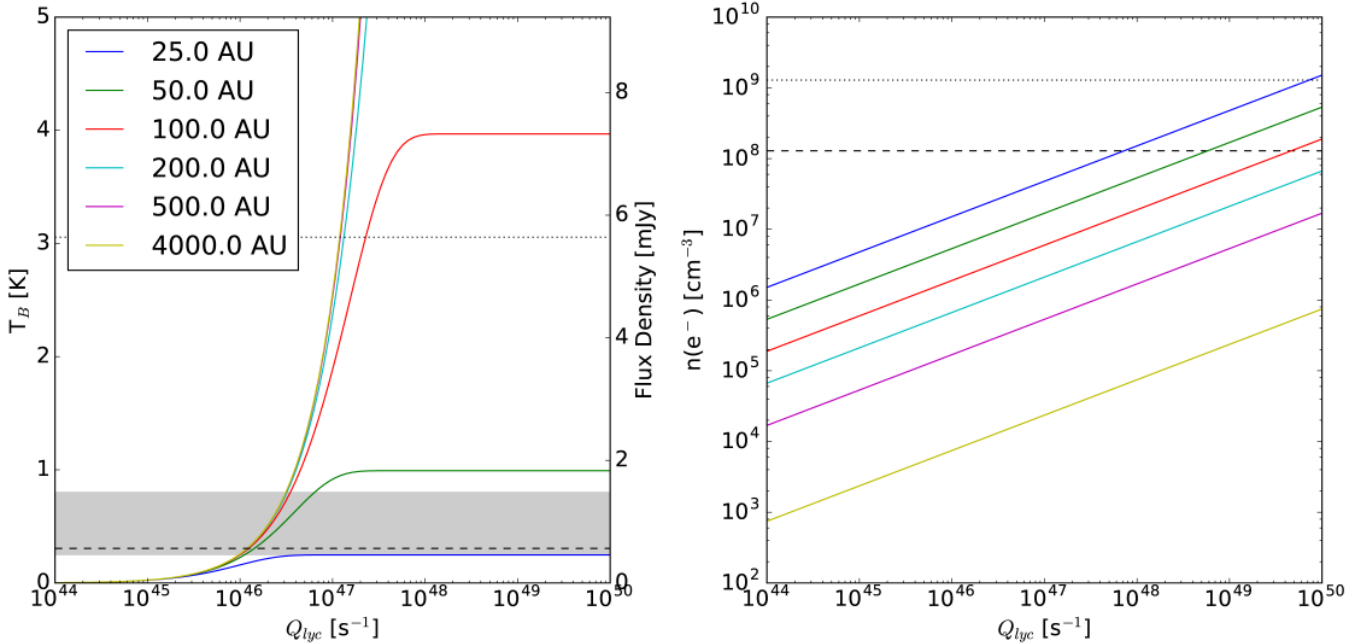
{Simple models of spherical \hii regions to illustrate the observable
properties of such regions.  The \hii region size is shown by line color; the
legend in the left plot applies to both figures.  (left) The expected
brightness temperature (left axis) and corresponding flux density at 95 GHz
within a FWHM=0.5\arcsec beam (right axis) as a function of the Lyman continuum
luminosity for a variety of source radii.  The grey filled region shows the
range of our 5-sigma sensitivity limits,
which vary with location from 0.25 to 0.8 K.
The dotted and dashed horizontal lines show the flux density of a 10 \msun
and 100 \msun isothermal dust core at $T=40$ K.
(right) The electron density required to produce an \hii region of radius
indicated by the legend in the left plot.  The
horizontal dashed line shows the density corresponding to an unresolved dust
source ($r<0.2\arcsec=1700$ AU) at the 5-$\sigma$ detection limit ($\approx0.5$
mJy, or 10 \msun of dust, assuming $T=40$ K, and assuming $n_e=2
n(\hh)$).    The dotted line shows the density corresponding to a 
100 \msun dust core at $T=40$ K.
}
{fig:hiibrightness}{1}{18cm}

\subsubsection{Alternative 5, our hypothesis: The sources are (mostly) YSOs}
\label{sec:theyareprotostars}
After determining that the other possibilities cannot explain
the whole sample, we test and validate the hypothesis
that most or all of the sources contain YSOs in this section.

If we assume the sources are dust-dominated and have a higher dust temperature
than used in Section \ref{sec:alt0}, the inferred gas mass is lower, but an
internal heating source - i.e., a protostar or young star - is required.  For
example, if we assume $T_D=80$ K\footnote{At these dust temperatures, we should
be concerned about the assumed opacity, since ices will begin to evaporate
\citep[e.g.,][]{Bergin1995a}, reducing the 3 mm opacity and correspondingly
increasing the required mass required to produce the observed flux
\citep{Ossenkopf1994a}.  }, our detection limit is only $M(80\mathrm{K}) = 4
\msun$.  Heating that much dust well above the cloud average requires a
high-luminosity central heating source.

To constrain the required heating source, we examine the protostellar models of
\citet[][specifically, the \texttt{spubhmi} and \texttt{spubsmi}
models]{Robitaille2017a} and \citet{Zhang2015f}.  The Robitaille models that
produce $S_{3 \mathrm{mm}} > 0.5$ mJy within an $r<2500\mathrm{~AU}$ aperture
uniformly have
${L>10^4~\lsun}$.  Such luminosities imply either that a high-mass
(${M\gtrsim8~\msun}$) star has already formed and is still surrounded by a
massive envelope or a high-mass YSO is present and accreting.
The models of \citet{Zhang2015f} generally only exhibit $L>10^4~\lsun$ once a
star has reached $M\approx10$ \msun as it continues to accrete to a higher
mass.  Similarly, pre-main-sequence stellar evolution models
\citep[e.g.,][]{Haemmerle2013a} only reach $L>10^4~\lsun$ at any point in their
evolution for stars with final mass $M\gtrsim8$ \msun.  In the
\citet{Robitaille2017a} model grid, all sources with $L>10^5~\lsun$ produce
$S_{3 \textrm{mm}}>0.5$ mJy, so our survey should be nearly complete to such
sources, but in the range $10^4 \lsun < L < 10^5 \lsun$, a substantial fraction
may be below our sensitivity limit.

\paragraph{Comparison to similar data}
We compare our detected sample to that of the Herschel Orion Protostar Survey
\citep[HOPS;][]{Furlan2016a} in order to get a general empirical sense of what
types of
sources we have detected.  We selected this survey for comparison because it is
one of the largest protostellar core samples with well-characterized bolometric
luminosities available.
Figure \ref{fig:hopshist} shows the HOPS source
flux densities at 870\um (from LABOCA on the APEX telescope) scaled to
$d=d_{Sgr B2}$ and 3 mm assuming a dust opacity index $\beta=1.5$,
which is shallower than usually inferred, so the extrapolated
fluxes may be slightly overestimated\footnote{\label{footnote:beta}
We err on the shallower side,
implying that the extrapolated 3 mm fluxes are brighter than they should be,
since this approach gives a more conservative view of the detectability of the
Orion sources.  In reality, such sources are likely even fainter than predicted
here.}.  The 870\um data
were acquired with a $\sim20\arcsec$ FWHM beam, which translates to a
resolution $\sim1\arcsec$ at $d_{Sgr B2} = $\dsgrb assuming $d_{Orion}=415$ pc,
so our beam size is somewhat smaller than theirs.

The HOPS sources are all fainter than even the faintest Sgr B2 sources.  The
most luminous
and brightest HOPS source, with $L_{tot}<2000$ \lsun, would only be 0.2 mJy in
Sgr B2, or about a 2-$\sigma$ source, which is below our detection threshold even in
the artifact-free regions of the map.  We  conclude that the Sgr B2 sources are
much more luminous than any in the Orion sample, which is consistent with all
of the sources in our sample being MYSOs.

\FigureOneCol{f13}
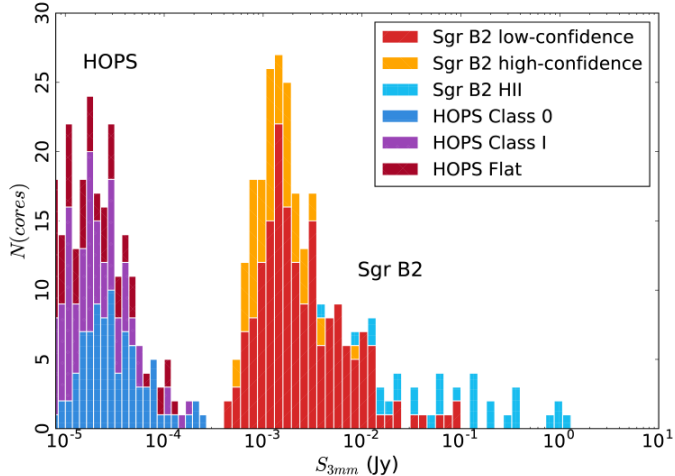
{A histogram combining the detected Sgr B2 cores with predicted flux densities
for sources at $d=\dsgrb$ and $\lambda=3$ mm
based on the HOPS \citep{Furlan2016a} survey.  The sources are labeled by their
infrared (2-20 \um) spectral index: Class 0 and I have positive spectral index
and flat spectrum sources have $-0.3 < \alpha_{IR} < 0.3$. The HOPS histogram
shows the 870 \um data from that survey scaled to 3 mm
assuming $\beta=1.5$ (see footnote \ref{footnote:beta}).
Every HOPS source is well below the detection threshold for our observations.}
{fig:hopshist}{1}{0.5\textwidth}

This conclusion is supported by a more direct comparison with the Orion nebula
as observed at 3 mm with MUSTANG \citep[][Figure
\ref{fig:orioncompare}]{Dicker2009a}.  Their data were taken at
9\arcsec FWHM resolution, corresponding to 0.48\arcsec at $d_{Sgr B2}$.  The
peak flux density measured in that map is toward Source I, $S_{90 GHz}(d_{Sgr
B2}) = 3.6$ mJy.  Source I\footnote{This source includes Source I, BN, and a few
other objects at this resolution, and at 3 mm Source I and BN are comparably
bright \citep{Plambeck2013a}.  This source is not part of the HOPS sample.}
would therefore  be detected and would be
somewhere in the middle of our sample.  It resides on a background of
extended emission, and the
extended component would be readily detected (and resolved) in our data. 
Source I is the only known high-mass YSO in the Orion cloud, and it would
be detectable in our survey, while no other compact sources in the Orion cloud
would be.  This comparison supports the interpretation that most of the
non-\hii region sources are massive YSOs.

\Figure{f14}
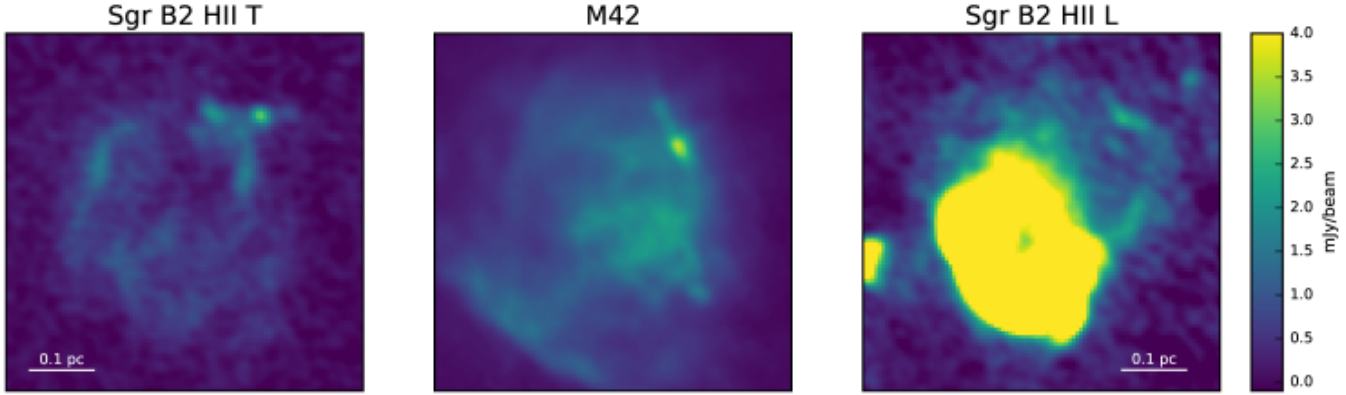
{Comparison of two extended \hii regions in Sgr B2 (ALMA 3 mm continuum) to the
M42 \citep[GBT MUSTANG 3 mm continuum;][]{Dicker2009a} nebula in Orion.
The three panels are shown on the same physical and color scale assuming
$d_{Orion} = 415$ pc and $d_{Sgr B2} = $\dsgrb and that the ALMA and MUSTANG
data have the same continuum bandpass.  Sgr B2 \hii T is comparable in
brightness and extent to M42; Sgr B2 \hii L is much brighter and is saturated
on the displayed brightness scale.  The compact source to the top right of the
M42 image is Orion Source I; the images demonstrate that Source I and the entire
M42 nebula would be easily detected in our data.
}
{fig:orioncompare}{1}{\textwidth}

\paragraph{The spectral indices of the dusty sources}
While we have concluded that the sources are dusty, massive YSOs, the
spectral indices we measured are somewhat surprising.  Typical dust clouds in
the Galactic disk have dust opacity indices $\beta\sim1.5-2$, implying 
a spectral index $\alpha\sim3.5-4$
\citep[$\beta=\alpha-2$;][]{Schnee2010a,Shirley2011a,Sadavoy2016a}.  Our spectral index measurements
are lower
than these:  only 3 sources out of \nalphas with significant $\alpha$
measurements have $\alpha > 3.5$\footnote{At the
$2\sigma$ level, up to 11 sources are consistent with $\alpha\geq3.5$, but this is
primarily because of their high measurement error.}, though \ngttwo of the
sources with $\alpha$ measurements have $\alpha>2$, indicating that their
emission is dust-dominated.  A shallower $\beta$ implies free-free
contamination, large dust grains, or optically thick surfaces are present
within our sources.  Since the arguments in previous sections suggest that the
sources are high-mass YSOs, the free-free contamination and optically thick
inner region models are both plausible.

\section{Analysis and discussion of star formation in Sgr B2}
\label{sec:sfdiscussion}

We have reported the detection of a large number of point sources and inferred
that they are most likely all high-mass YSOs.  In this section,
we discuss
the source flux density distribution function and star formation rate estimates
(\S \ref{sec:distributionsandsfr}), the difference between the clustered and
distributed source populations (\S \ref{sec:clustersandextended}), star
formation surface density thresholds (\S \ref{sec:thresholds}), star formation
and gas surface density relations (\S \ref{sec:gutermuth}), and the
implications of a varying volume density threshold (\S \ref{sec:turbsftheory}).

\subsection{Source distribution functions and the star formation rate}
\label{sec:distributionsandsfr}

In this section we examine the distribution of observed flux densities and the
implied total stellar masses.

If we make the very simplistic, but justified (Section
\ref{sec:theyareprotostars}), assumption that the sources we detect all contain
YSOs with $L_{bol}\gtrsim10^4$ \lsun, and in turn make the related
assumption that each source either currently contains or will form into an
$M\gtrsim8 \msun$ star, we can infer the total (proto)stellar mass in the
observed region.

We assume the stellar masses based on the arguments in Section
\ref{sec:theyareprotostars}: in order to be detected, the sources must either
be active OB stars illuminating \hii regions, very compact cores with $M>10$
\msun of warm dust within $R<4000$ AU, or at least moderately-massive YSOs
within warm envelopes.  Note that the mass estimates in this section are for
the resulting stars, not their envelopes.

\FigureOneCol{f15}
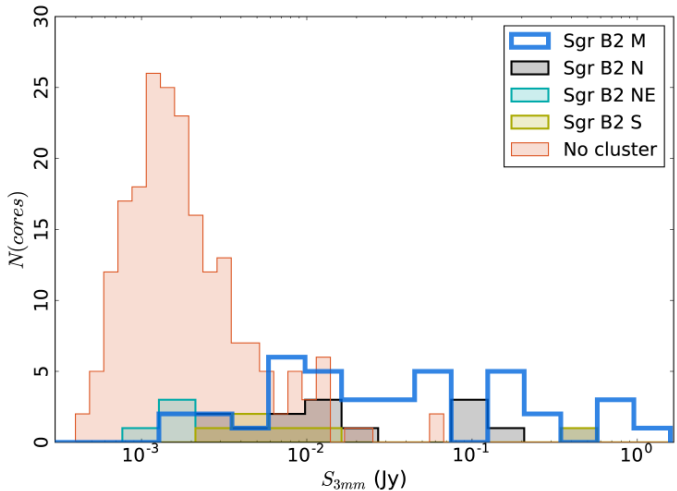
{Histograms showing the flux density (the peak intensity converted to flux density
assuming the source is unresolved) of the observed sources classified by their
cluster association.  Unlike Figure \ref{fig:fluxhist}, the histograms are
overlapping, not stacked.  The bin widths for the clusters are wider than
for the unassociated sources.}
{fig:fluxhistclusters}{1}{0.5\textwidth}

To compute the total mass of the forming star populations,
we assume each source not associated with an \hii region contains or will form
a star with mass equal to the average over the range 8-20 \msun assuming a
\citet[][Eqn. 2]{Kroupa2001a} initial
mass function, $\bar{M}(8-20) = 12~\msun$ (in this section, we refer
to these objects as ``cores'').  Based on the arguments in Section
\ref{sec:theyarehiiregions}, we assume each \hii region contains a star that is
B0 or earlier, and therefore that they each have a mass equal to the
average over 20 \msun, $\bar{M}(>20) = 45~\msun$.  In Table
\ref{tab:clustermassestimates}, the total counted mass estimate is shown as
$M_{count} = N \bar{M}$, where $N$ is the number of stars with an assumed mass
$\bar{M}$.

We also compute the \emph{total} stellar mass (i.e., the extrapolated mass
including low-mass stars) using the mass fractions $f(M>20) = 0.14$ and
$f(8<M<20)=0.09$ derived from the assumed IMF.   The total mass inferred by
extrapolating our measurements with this IMF is then 
\begin{subequations}
\begin{align}
    M_{\mathrm{inferred,\hii}}  & = M_{\mathrm{count}}(M>20) / f(M>20) \\
    M_{\mathrm{inferred,cores}} & = M_{\mathrm{count}}(8<M<20) / f(8<M<20) \\
    M_{\mathrm{inferred}}       & = (M_{\mathrm{inferred,cores}} + M_{\mathrm{inferred,\hii}})/2 \\
                                & = M_{\mathrm{count}}(M>8) / f(M>8) 
\end{align}
\end{subequations}
The inferred masses computed from \hii region
counts and from core counts are shown in columns $M_{inferred,\hii}$ and
$M_{inferred,cores}$ of Table~\ref{tab:clustermassestimates} respectively.
$M_{inferred}$ is the average of these two
estimates; it is also what would be obtained if all stars were assumed to be
average stars with $M>8$ \msun.  If our mass range classifications are correct
and the mass distribution is governed by a power-law IMF, we expect
$M_{inferred,\hii} = M_{inferred,cores}$.

We identify each source as belonging to one of the clusters described in
\citet[][see Figure \ref{fig:overview}]{Schmiedeke2016a}. In each cluster, we
count the number of \hii regions identified in our survey plus those identified
in previous works \citep{Gaume1995a,De-Pree1996a}, and we count the number of
protostellar cores not associated with \hii regions.  The distributions of
source flux densities associated with each cluster are shown in Figure
\ref{fig:fluxhistclusters}.  The cluster affiliation for each source is
reported in Table~\ref{tab:photometry}.

In Sgr B2 N and S, the core-based
and \hii-region based estimates agree to within a factor of 2, which is about
as good as expected from Poisson noise in the counting statistics.  
Sgr B2 M contains the largest source sample, and it has a factor of nine
discrepancy between the core and \hii-region based counts. The discrepancy
may arise from the combined effects of source
confusion at our 0.5\arcsec resolution and the increased noise around the
extremely bright central region that makes detection of $<2$ mJy sources
difficult.  The majority of pixels within the cluster region have significant
detections at 3 mm, but we do not presently have the capability to distinguish
between extended dust emission, free-free emission, or a confusion-limited
point source population.  While it is possible that this discrepancy
is driven by observational limitations, we also explore in Section
\ref{sec:clustersandextended} the possibility that it is a real physical
effect.

We compare our mass estimates to those of \citet{Schmiedeke2016a}, who inferred
stellar masses from \hii region counts.  The  two columns of Table
\ref{tab:clustermassestimates} with superscript $S$ show their observed and
estimated masses based on
\hii region counts.  For Sgr B2 M and N, our results are similar, as expected
since our catalogs are similar.  For S and NE, we differ by a large factor,
primarily because \citet{Schmiedeke2016a} assumed that $M_{min,YSO}$ and $M_{max}$
were the smallest and largest observed masses in the cluster, while we assumed
$M_{min,MYSO}=8$ \msun and $M_{max}=200$ \msun; i.e., we assumed a spatially
invariant IMF, while they assumed their observed sources represent a smaller
fraction of the integrated IMF and therefore their assumed mass fraction is
less than ours; $f(M_{min}< M < M_{max}) < f(M>20)$.

\begin{table*}[htp]
\centering
\caption{Cluster Masses}
\begin{tabular}{cccccccccc}
\label{tab:clustermassestimates}
Name & $N(cores)$ & $N(H\textsc{ii})$ & $M_{count}$ & $M_{inferred}$ & $M_{inferred, H\textsc{ii}}$ & $M_{inferred, cores}$ & $M_{count}^s$ & $M_{inf}^s$ & SFR \\
 &  &  & $\mathrm{M_{\odot}}$ & $\mathrm{M_{\odot}}$ & $\mathrm{M_{\odot}}$ & $\mathrm{M_{\odot}}$ & $\mathrm{M_{\odot}}$ & $\mathrm{M_{\odot}}$ & $\mathrm{M_{\odot}\,yr^{-1}}$ \\
\hline
M & 17 & 47 & 2300 & 8800 & 15000 & 2300 & 1295 & 20700 & 0.012 \\
N & 11 & 3 & 270 & 1200 & 980 & 1500 & 150 & 2400 & 0.0017 \\
NE & 4 & 0 & 48 & 270 & 0 & 540 & 52 & 1200 & 0.00037 \\
S & 5 & 1 & 110 & 500 & 330 & 680 & 50 & 1100 & 0.00068 \\
Unassociated & 203 & 6 & 2700 & 15000 & 2000 & 27000 & - & - & 0.02 \\
Total & 240 & 57 & 5500 & 26000 & 19000 & 33000 & 1993 & 33400 & 0.035 \\
Total$_{max}$ & - & - & - & 46000 & - & - & - & - & 0.062 \\
\hline
\end{tabular}
\par
$M_{count}$ is the mass of directly counted protostars, assuming each millimeter source is 12.0 \msun, or 45.5 \msun if it is also an \hii region.  $M_{inferred,cores}$ and $M_{inferred,\hii}$ are the inferred total stellar masses assuming the counted objects represent fractions of the total mass 0.09 (cores) and 0.14 (\hii regions).  $M_{inferred}$ is the average of these two.  $M_{count}^s$ and $M_{inf}^s$ are the counted and inferred masses reported in \citet{Schmiedeke2016a}.  The star formation rate is computed using $M_{inferred}$ and an age $t=0.74$ Myr, which is the time of the last pericenter passage in the \citet{Kruijssen2015a} model.  The \emph{Total} column represents the total over the whole observed region.  The \emph{Total}$_{max}$ column takes the higher of $M_{inferred,\hii}$ and $M_{inferred,cores}$ from each row and sums them.  We have included \hii regions in the $N(\hii)$ counts  that are \emph{not} included in our source table \ref{tab:photometry} because they are too diffuse, or because they are unresolved in our data but were resolved in the \citet{De-Pree2014a} VLA data.  As a result, the total source count is greater than the source count reported in Table \ref{tab:photometry}. Also, the unassociated \hii region count is incomplete; it is missing both diffuse \hii regions and possibly unresolved hypercompact \hii regions, since there are no VLA observations comparable to \citet{De-Pree2014a} in the unassociated regions.
\end{table*}

\subsubsection{Sgr B2's star formation rate}
\label{sec:sfr}
We estimate the star formation rate using the above mass estimates.  To
determine the star formation rate, we need to know the age of the current star
forming burst.  We use the dynamical model of \citet{Kruijssen2015a} to get an
age of the Sgr B2 cloud $t=0.74$ Myr, the time since
pericenter passage.  We divide the inferred stellar mass by this
age\footnote{We use the higher of the two masses out of $M_{inferred,\hii}$ and
$M_{inferred,cores}$ for each row because, as discussed in Section
\ref{sec:clustersandextended}, Sgr B2 M likely has an underestimated
$M_{inferred,cores}$ either due to observational effects such as confusion or
because it is older and the more moderate-mass sources represented by the cores
have become unobservable.  Similarly, the unassociated sources appear
to be younger and therefore the \hii-region based mass appears
to be an underestimate.}; the results are shown in
Table~\ref{tab:clustermassestimates}.    Our estimated total inferred SFR of
the Sgr B2 cloud is \sfrdynagemax \msun \peryr,  at least half of the
total for the CMZ \citep[$\dot{M}_{CMZ}=0.07-0.12$ \msun
\peryr;][]{Longmore2013b,Barnes2017b}.  

However, there are several assumptions that go into the above calculations:
\begin{itemize}
    \item The computed rate assumes that star formation was initiated at the
        cloud's most recent pericenter passage following the
        \citet{Kruijssen2015a} orbital model.  Other models for the CMZ
        dense gas have been discussed
        \citep[e.g.,][]{Molinari2011a,Sofue2017b,Ridley2017a,Sormani2017a},
        though \citet{Henshaw2016a} found that the \citet{Kruijssen2015a} model
        best fit the data.
    \item In the context of the Kruijssen et al model, we have used the time
        since pericenter passage as $t_{sf}$, but G0.253+0.016 shows almost no
        star formation; the appropriate timescale may instead be the time since
        Sgr B2 was at the position of G0.253, approximately $t_{sf}=0.43$ Myr.
        This shorter age would yield a SFR
        ${\dot{M}=\sfrbrickagemax~\msun~\peryr}$, which would imply that Sgr
        B2 completely dominates the instantaneous SFR of the CMZ.
    \item It  assumes that all stars whose passage was triggered at that event
        are still visible as 3 mm cores to our survey, but it is possible that
        the lifetime of these cores is shorter than 0.74 Myr.  For example,
        low-mass Class 0 cores have lifetimes 0.16 Myr and Class I have
        lifetimes 0.54 Myr \citep{Evans2009a}.  If we are only sensitive to 
        more massive analogues of Class 0 sources, many of the already formed
        stars will have become undetectable, resulting in our rate being
        an underestimate.  Section \ref{sec:theyareprotostars} argues
        they are probably a mix of Class 0 and I equivalent sources,
        but the lifetimes of the massive analogues are unconstrained
        and could be shorter.
\end{itemize}

While our measurements of the total star formation activity in Sgr B2 are
likely the best to date, our estimate of the star formation rate remains
strongly dependent on the assumed star formation timescale.

\subsection{The clusters and the extended population}
\label{sec:clustersandextended}
We noted in Section \ref{sec:distributionsandsfr} that the \hii-region-inferred
protostellar mass matches the core-inferred protostellar mass to within a
factor of 2 in the whole Sgr B2 cloud and the individual clusters excepting Sgr
B2 M.  In Sgr B2 M, the \hii-region inferred mass is $\sim9\times$ greater than
the core-inferred mass.  While the lack of faint sources in Sgr B2 M could
be an observational limitation, it may be a real effect signifying an evolutionary
difference.

Sgr B2 M has more \hii regions and is more centrally condensed than any of the
other clusters and the distributed star forming population.  Assuming that \hii
regions represent a later stage in protostellar evolution than the dusty
protostellar core stage, the \hii region excess in Sgr B2 M implies that it is
older than Sgr B2 N and the distributed protostar population.  By contrast,
along the Sgr B2 DS ridge, there are no \hii regions, but there are $\sim100$
high-mass YSOs, which implies that these YSOs began their formation
nearly simultaneously.  Figure \ref{fig:fluxhistclusters} shows this difference
graphically; Sgr B2 M has an overall source flux distribution marginally higher
than Sgr B2 N but dramatically higher than the unclustered sources.

The large number of probable YSOs observed along an elongated ridge allows us
to estimate an upper limit on their age.  Assuming all of these forming stars
are bound to the cloud and/or central clusters, they should approach a
spherical distribution within about one crossing time \citep{Efremov1998a}.  If
we assume the turbulent velocity dispersion is $\sigma_{1D}\approx10$ \kms
\citep[e.g.,][]{Henshaw2016a}, and the length of the DS ridge is $L\approx10$
pc, the upper limit on the formation time of the YSOs is $L/\sigma_{1D}<1$ Myr.
Most of the sources along the ridge are within $r<0.5$ pc of it center (Figure
\ref{fig:coreson20cmandhc3n}),
which, assuming they
formed in the ridge, suggests an upper age limit $t<r/\sigma_{1D}=5\ee{4}$ yr
\citep[however, the stars may have a lower velocity dispersion by a factor of
5-10, implying a more conservative upper age limit is $t<0.5$
Myr; ][]{Offner2009c}.
The DS ridge sources appear to be recently formed, which may explain the
relative lack of \hii regions in the distributed population (Table
\ref{tab:clustermassestimates}): the forming massive stars have not yet had
time to contract and produce ionizing radiation.

The expanding \hii regions observed around Sgr B2 M and N (and assumed to be
associated with them) give a lower limit on their ages \citep[assuming steady
expansion, which may not be a correct model;][]{Peters2010b,De-Pree2014a}.  The
\hii regions I, J, A1, and K4 have radii $r\approx0.1$ pc \citep{Gaume1995a},
suggesting their ages are at least $t>10^5$ yr assuming they are expanding into
a density $n\gtrsim10^5$ \percc \citep{De-Pree1995a,Schmiedeke2016a}.  The
clusters therefore appear to be somewhat older than the ridge sources.

The relative ages of M and the rest of the region (i.e., Sgr B2 M is apparently
older) suggest two possibilities for their formation history.  If we take the
ages at face value, Sgr B2 M must have collapsed first to form stars in an
early event, then the DS ridge began forming stars in a
subsequent event.  A second possibility is that the overall collapse of both Sgr B2
M and DS began at the same time, but the Sgr B2 M region was denser and
therefore had a shorter collapse time, which is predicted by hierarchical
cluster formation models to lead to higher star formation efficiencies
\citep{Kruijssen2012a}.  Our catalog does not allow us to distinguish these
possibilities.  However, the latter scenario would predict that the cloud
should be in a state of global collapse, with the least dense regions
collapsing most slowly.  This collapse has been suggested to be ongoing in CMZ
clouds by \citet{Walker2015a,Walker2016a} and may leave detectable kinematic
signatures (e.g., self-absorption in moderately optically thick lines) in the
dense gas.

\citet{Yusef-Zadeh2009a} noted the presence of some Spitzer 4.5 \um excess
sources and 24 \um sources in the southern part of Sgr B2, and from these
detections concluded that star formation had proceeded outside-in in the Sgr B2
cloud.  Our data have revealed a much larger population of what are most likely
younger sources (dust-dominated YSOs) in this region, which is
inconsistent with the previous interpretation.  Instead, it seems that the
central clusters are the oldest sites of star formation.  The excess of 4.5 \um
and 24 \um sources in DS may be because the cloud's envelope of opaque material
is thinner along those lines-of-sight.  We conclude that existing infrared
observations of the Sgr B2 cloud lack both the depth and resolution to detect
the significant ongoing star formation we report here.

\subsection{An examination of star formation thresholds}
\label{sec:thresholds}
Several authors \citep[e.g.,][]{Lada2010a,Heiderman2010a} have proposed that star
formation can only occur above a certain density or column density
threshold\footnote{Column density is commonly used as a proxy for volume
density because of its observational convenience, but volume density is the
more meaningful physical parameter for most relevant processes in star formation
(e.g., gravity and pressure).}.  \citet{Kruijssen2014c} suggested that the
column density threshold in the CMZ should be higher than that in local clouds
based on predictions from turbulence-based star formation theories
\citep{Krumholz2005c,Padoan2011b}.
We therefore discuss our measurements of column density thresholds in this
section.

\subsubsection{Comparison to the Lada, Lombardi, and Alves 2010 column density threshold}
\label{sec:ladathreshold}
In this section, we compare the star formation threshold in Sgr B2 to that in
local clouds performed by \citet[][hereafter, LLA10]{Lada2010a}.  They determined
that all star
formation in local clouds occurs above a column density threshold $M_{thresh} >
116$ \msun pc$^{-2}$, or $N_{thresh}(\hh) > 5.2\ee{21}$ \persc assuming the
mean particle mass is 2.8 amu \citep{Kauffmann2008a}.  We first note, then,
that \emph{all pixels} in our column density maps (Section \ref{sec:colmaps},
Battersby et al, in prep) are above this threshold by \emph{at least} a factor
of 10.

LLA10 identified their star-formation threshold by comparing the
cumulative column density to total YSO count across a range of clouds and
identifying the point of minimum variance.  Our sample covers only one cloud,
so we cannot perform the same analysis.  Instead, we examine the column density
above which high-mass YSOs (`Class 0/I'-like sources, since they have dust
envelopes) are forming.

Figure \ref{fig:corebackgroundcdf} shows the cumulative distribution function
of the column density associated with each identified continuum source; the
column density used is the nearest-neighbor pixel to the source in the column
density maps.  Even using the conservative maximum temperature $T_{dust}=50$ K
(resulting in the minimum column density), all of the sources exist at a column
density an order of magnitude higher than the Lada threshold, and they exist
above that threshold even if the foreground is assumed to be $5\ee{22}$ \persc,
the highest plausible value considered in Section \ref{sec:colmaps}.
While all of the sources exist above the Lada threshold, not all pixels above
this threshold contain YSOs or protostellar cores (Figure
\ref{fig:coldistributionwithoutstarformation}).

\FigureOneCol{f16}
{
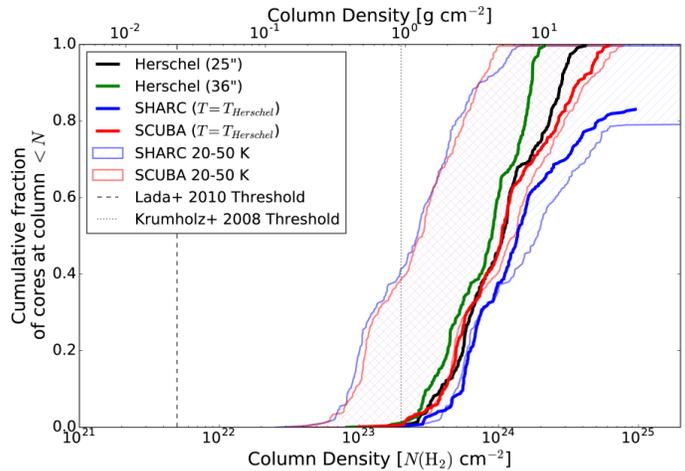
Cumulative distribution functions of the background column density associated
with each identified 3 mm continuum source.  The column densities are computed
from a variety of maps with different resolution and assumed temperature.
The Herschel maps use SED-fitted temperatures (Battersby et al. in prep) at
25\arcsec resolution (excluding the 500 \um data point) and 36\arcsec resolution.
The SHARC 350 \um and SCUBA 450 \um maps both have higher resolution ($\sim10\arcsec$)
but no temperature information; we used an assumed $T_{dust}=20$ K and $T_{dust}=50$ K
to illustrate the range of possible background column densities (hatched
red and blue).  The thick solid red and blue lines show the SHARC and SCUBA column
density images using Herschel temperatures interpolated onto their grids: these
curves are closer to the 20 K than the 50 K curve and serve as the best estimate
column density maps.  The SHARC data fail to go to a cumulative fraction of 1
because the central pixels around Sgr B2 M and N are saturated (the lower temperature
assumptions result in optical depths $>1$, which cannot be converted to column
densities using the optically thin assumption).  The vertical
dashed line shows the $N(\hh)=5.2\ee{21}$ \persc column density threshold from
LLA10, and the vertical dotted line shows the the $N(\hh)=2\ee{23}$
\persc \citet{Krumholz2008a} threshold for high-mass star formation.}
{fig:corebackgroundcdf}{1}{0.5\textwidth}

LLA10 suggested that their observed column density threshold
corresponds to a density $n\approx10^4$ \percc.  If we assume that the dense
part of the  Sgr B2 cloud is
approximately a $2 \mathrm{pc}\times2 \mathrm{pc}\times6 \mathrm{pc}$ box
(i.e., we assume the depth is equal to the shortest observed dimension on the
sky), the typical column density $N\gtrsim5\ee{23}$ \persc requires a mean
density $n\gtrsim5\ee{4}$ \percc (which is a lower limit; most of the mass
is at higher column densities).  Again, effectively all of the gas associated
with ongoing star formation is above the locally-determined threshold.

To compare Sgr B2 to the LLA10 sample on a full-cloud scale, we can use
the total cloud mass and total YSO mass.  LLA10 used a YSO count, $N_{YSO}$,
while we infer a total YSO mass; we use their assumed median mass $M_{median} =
0.5$ \msun to convert our observed $M_{YSO}$ to $N_{YSO}$.  Using their
fitted relation for local clouds\footnote{In the main body of their paper, Lada
et al included all YSOs in the clouds down to $A_K>0.1$ for their total YSO
counts.  However, in the text they repeated their $N_{YSO}$ - $M_{cloud}$ fit
using only stars embedded in gas with $A_K>0.5$.  They obtained a linear
relation about 0.6$\times$ lower than that shown in their paper (C. Lada,
private communication).  The better agreement when including only embedded
YSOs hints that the discrepancy noted in this section could disappear if a
complete census of Class II sources were obtained in Sgr B2.}, 
$N_{YSO,Lada} = 0.2 M_{cloud,\msun}(A_K>0.8)$,
we predict that for $M_{Sgr B2}=1.5\ee{6}$ \msun (where we use the whole cloud
mass because all of the cloud is at $A_K>0.8$), $N_{YSO,Sgr B2,Lada} =
3\ee{5}$.  As seen in Table~\ref{tab:clustermassestimates}, the observed
$N_{YSO,Sgr B2,obs} = M_{YSO} /
(0.5 \msun) = 5.2\ee{4}-9.2\ee{4}$ \msun, a factor of three to six below the
extrapolated LLA10 relation.  If we invert the equation to obtain a cloud
mass from our
observed $N_{YSO,Sgr B2,obs}$, we would predict $M_{cloud,Lada}\approx2.6-4.6\ee{5}$ 
\msun,
which is equivalent to the mass in Sgr B2 above a column density threshold
$N>0.8-1\ee{24}~\persc$ (Figure \ref{fig:cumulativemass}).

Any way we examine our data, it appears that a higher column density threshold
for star formation is required in Sgr B2 than in local clouds.  The one
remaining caveat is that the LLA10 study  used Spitzer observations of nearby
clouds that were nearly complete to stars at least as small as 0.5 \msun.  By
contrast, as discussed in Section \ref{sec:theyareprotostars}, our survey is
sensitive only to stars with $M\gtrsim8$ \msun.  It is therefore possible that
we have instead observed a higher threshold for high-mass star formation than
for low-mass star formation \citep[e.g., as suggested by ][]{Krumholz2008a}.

\FigureOneCol{f17}
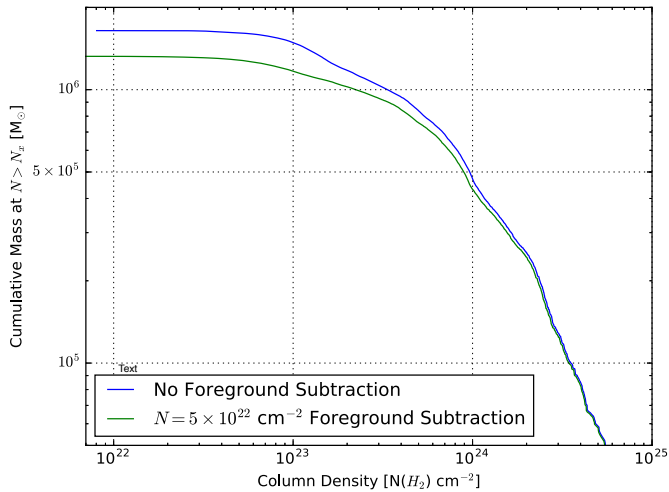
{The cumulative mass above a threshold column density in the observed
region.  The two curves show the mass inferred with and without a
foreground of $5\ee{22}$ \persc, the highest plausible foreground column
density, subtracted.}
{fig:cumulativemass}{1}{0.5\textwidth}

\subsubsection{Other Thresholds}
\label{sec:otherthresholds}

A theoretical threshold for high-mass star formation, $\Sigma > 1$ g \persc
($N(\hh) > 2\ee{23}$ \persc) was developed by \citet{Krumholz2008a}.   Nearly
all of the sources we have detected reside above this threshold (independent of
the assumed foreground contamination), and we determined our sources are all
likely to be massive YSOs in Section \ref{sec:theyareprotostars}.
However, not all pixels with $\Sigma > 1$ g \persc are forming high-mass stars
(Figure \ref{fig:coldistributionwithoutstarformation}).  It appears
there is a threshold, but it is a necessary, not a sufficient, criterion
for high-mass star formation.

However, there is another threshold in our data, $N(\hh)>1\ee{24}$ \persc,
above which the majority of the gas is associated with ongoing high-mass star
formation (Figure \ref{fig:coldistributionwithoutstarformation}).  This
threshold suggests that any gas reaching a column density $N(\hh)>10^{24}$
\persc over a $\approx0.5$ pc size scale (the resolution of our column density
maps) has more likely begun to form high-mass stars.  This column
density corresponds to a volume density $n(\hh)\approx10^5$ \percc assuming
spherical symmetry.

\FigureOneCol{f18}
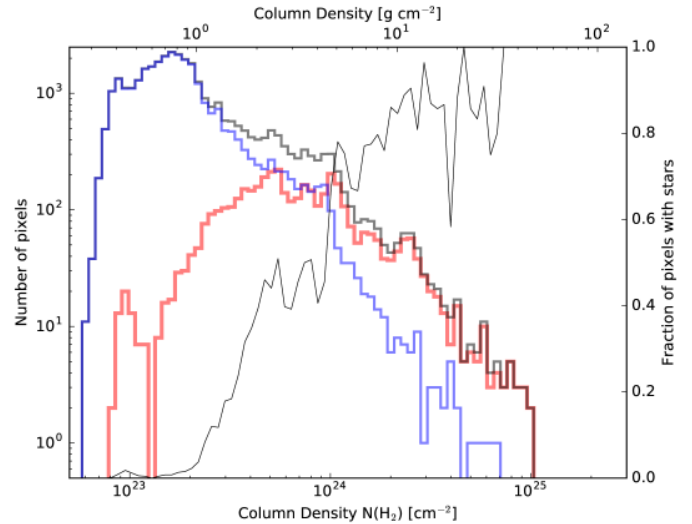
{Histograms of the column density measured with the combined SCUBA and Herschel
data using the interpolated Herschel temperatures covering only the region
observed with ALMA.  The black histogram (left axis) shows the whole observed region,
the blue solid line shows the SCUBA pixels that do not contain an ALMA source,
and the red thick line shows those pixels that are within one beam
FWHM of an ALMA source.  The thin black line (right axis) shows the ratio of
the red histogram to the black histogram, i.e., it shows the fraction of pixels
with associated
YSOs.  While the ALMA sources (high mass YSOs)
clearly reside in high-column gas, there is abundant high-column-density
material that shows no signs of ongoing star formation.}
{fig:coldistributionwithoutstarformation}{1}{0.5\textwidth}

\subsubsection{Comparison to G0.253+0.016}
In G0.253+0.016 (The Brick, G0.253), very little star formation
has been observed
\citep{Longmore2013a,Johnston2014a,Rathborne2014a,Rathborne2015a} despite most
of the cloud existing above the locally measured LLA10 column
density threshold.  The column density distribution for G0.253
is shown in Figure \ref{fig:bricksgrb2colcompare}.

The \citet{Rathborne2014a} and \citet{Rathborne2015a} ALMA 3 mm data are the
deepest observations of G0.253 in the millimeter regime to date, with a
sensitivity about $4\times$ better than ours, but a beam of 1.7\arcsec (similar
to that shown in Figure \ref{fig:overview}; compare to Figure 2 in both
Rathborne et al papers).  Despite the higher sensitivity of their data,
they detected only 3 compact continuum sources.  Similarly,
\citet{Kauffmann2013a} detected only one compact continuum source in their
(less sensitive) SMA data.  By contrast, even in our coarse resolution data, which
have a worse sensitivity (RMS $\approx 0.25$ mJy beam$^{-1}$, $10\times$ worse
than Rathborne et al), dozens of compact sources are evident.  Our better
resolution was critical for identifying the hundreds of sources we have
identified, but it is nonetheless clear that the star formation activity is
much higher in Sgr B2 than G0.253.

Comparing Sgr B2 to G0.253, the majority of the Sgr B2 cloud is at higher
column than G0.253.  Star formation in Sgr B2 nearly all occurs
at a higher column than exists within G0.253 (Figure
\ref{fig:bricksgrb2colcompare}).  The dearth of observed cores in G0.253 is
therefore easily explained if there is a density threshold for star
formation that is not reached in G0.253.  Given that the G0.253 observations
were deeper than our own, yet still identified almost no forming stars, it
appears more likely that there is a lack of star formation rather than simply a
lack of high-mass star formation.  Nonetheless, robust verification of this
hypothesis will require much deeper observations sensitive to low-mass stars
in both regions.

\FigureOneCol{f19}
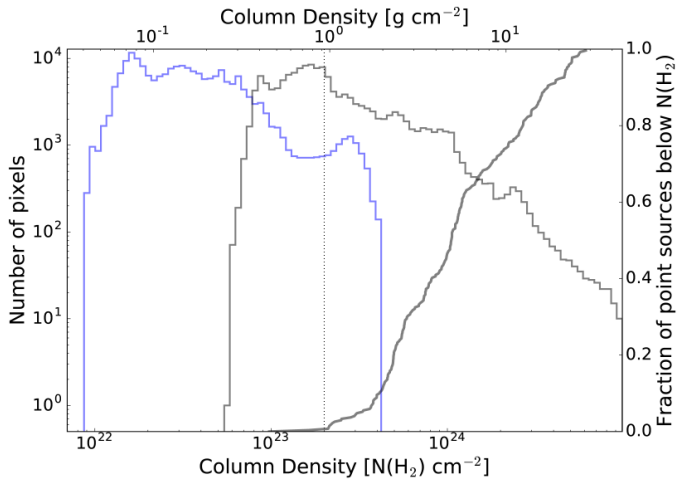
{Histograms of the column density of G0.253+0.016 (blue) and Sgr B2 (gray)
using the combined SCUBA 450 \um and Herschel 500 \um intensity with the
interpolated Herschel dust temperatures.  The cumulative distribution of core
`background' column densities in Sgr B2 is shown as a thick gray line, showing
that the densities at which stars are forming in Sgr B2 are barely
reached in G0.253.  The vertical dotted line is the \citet{Krumholz2008a}
threshold for high-mass star formation at $N(\hh)=2\ee{23}$ \persc, while
the \citet{Lada2010a} threshold is below the minimum value plotted here (see
Section \ref{sec:thresholds}).}
{fig:bricksgrb2colcompare}{1}{0.5\textwidth}

\subsection{Surface density relations: comparison to Gutermuth et al. 2011}
\label{sec:gutermuth}
Unlike \citet{Lada2010a}, who invoke a threshold followed by a linear star
formation law relating the gas to the stellar surface density,
\citet{Gutermuth2011a} concluded that star formation was best represented as
power-law relations between the stellar and gas mass surface densities.

In this section, we measure the stellar surface density (\S
\ref{sec:gutermuthmethods}) and compare the star-gas surface density relation
to the local clouds observed by \citet[][\S
\ref{sec:gutermuthcomparison}]{Gutermuth2011a}, finding that the local clouds
and Sgr B2 do not fit on a common relation.  We examine the possible reasons
for the disagreement (\S \ref{sec:gutermuthdiscrepancy}), concluding that a
varying volume density threshold for star formation is the most likely
explanation.

\subsubsection{Methodological comparison to Gutermuth et al}
\label{sec:gutermuthmethods}
We adopt the same approach used in \citet{Gutermuth2009a} and
\citet{Gutermuth2011a} to compare gas and stellar mass surface densities.  We
computed both the star-centric mass surface density using the 11th nearest
neighbor density and a gridded surface density.  We assume a mean mass per
source $\bar{M}(M>8~\msun)=21.8$ \msun, and that each such star represents 23\%
of the total stellar mass (see Section \ref{sec:distributionsandsfr}), i.e.,
each 3 mm source is treated as a ``cluster'' containing 95 \msun of stellar
mass\footnote{In previous
sections, we assigned different masses to different source classes, i.e., we
assigned higher masses to \hii regions than non-\hii regions.  For consistency
with \citet{Gutermuth2011a}, we assume a constant mass per source here, which
may result in a systematic underestimation of the stellar mass surface density
at the highest densities (since the \hii regions are preferentially
concentrated in clusters).}. The correlation is similar whether we use the
Herschel column density directly or the SCUBA or SHARC-based column density
maps (see Section \ref{sec:colmaps}).  

There are a few key differences between our data and those of
\citet{Gutermuth2011a}.  First, our \emph{minimum} detected column density is
$N(\hh)\approx10^{23}$ \persc, while in their sample, the \emph{maximum}
observed was $A_V=38$, or $N(\hh)=3.8\ee{22}$ \persc.  Even if we subtract our
upper-limit foreground estimate $N(\hh)=5\ee{22}$ \persc from the entire Sgr B2
map, nearly all of the detected sources reside in regions with column densities
well above the maximum reached in the local cloud sample.  Second, our 3 mm
source sample is sensitive to only the youngest sources, either the high-mass
equivalent of Class 0/I sources (`hot cores' or HMYSOs), or deeply embedded
hypercompact \hii regions.  The Spitzer sample included both Class~I sources,
with estimated ages $t\lesssim0.5$ Myr, and Class~II sources, with ages $0.5 <
t < 5$ Myr.  Our sample is therefore biased young compared to theirs.  If the
age estimate for Sgr B2 from the dynamical models
\citep{Kruijssen2015a} is accurate, there should be about as many
Class II sources as Class I, given the standard ages, meaning our total stellar
mass estimate may be as much as a factor of 2 underestimated.  Third, as noted
above, we are sensitive to only high-mass sources, so we infer a significant
population that is not directly observed.

We computed star formation relations following \citet{Gutermuth2011a} Section 4.1.
We use their equation 7:
\begin{equation}
    \label{eqn:stellarsurfdens}
    \Sigma_*(t) = c \Sigma_{gas,0} \left[1-\left(\frac{t}{t_0}+1\right)^\beta\right]
\end{equation}
where $\Sigma_*(t)$ is the time-dependent stellar surface density, $c$ is a
scaling constant (assumed to be the star formation efficiency of a core and
to have the value 0.3), $\Sigma_{gas,0}$ is the initial gas surface density,  $t_0$
is the timescale for the gas to be depleted by $2^{\beta}$, $\beta =
1/(1-\alpha)$, and $\alpha$ is
the exponent in the star formation relation ($\alpha\ne 0$).
The depletion timescale $t_0$ is defined by their equation 5:
\begin{equation}
    t_0 = \frac{1}{k(\alpha-1)} \Sigma_{gas,0}^{1-\alpha}
\end{equation}
where $k$ is the star formation rate coefficient.  The constant $k$ has
different units depending on which value of $\alpha$ is adopted;
for $\alpha=2$, $k$ has units pc$^{2}$ \msun$^{-1}$ Myr$^{-1}$.

If $\alpha=1$, i.e., the star formation rate is proportional to the intial gas
surface density, the surface density relation is instead $\Sigma_*(t) = c
\Sigma_{gas,0}[1-e^{-kt}]$ and the 50\% depletion timescale is $t_0=\ln(2)/k$.
The constant $k$ is then the inverse star formation timescale with units
Myr$^{-1}$

\subsubsection{Results of the comparison to Gutermuth et al}
\label{sec:gutermuthcomparison}

Figure \ref{fig:stellarvsmasscolumn} shows the stellar mass surface density
$\Sigma_*$ plotted against the gas mass surface density $\Sigma_{gas}$.
Our data show a large scatter and are plausibly compatible with a
power-law index in the range 1-2, and therefore may be consistent with the
steep slopes ($\alpha\approx2$) \citet{Gutermuth2011a} derived.
\citet{Lada2017a} and \citet{Lombardi2014a} derived similarly steep
slopes ($\alpha=2$ for Orion, $\alpha=3.3$ for the California cloud;
see Appendix \ref{sec:ladasurfdensappendix}).

\FigureTwo
{f20}
{f21}
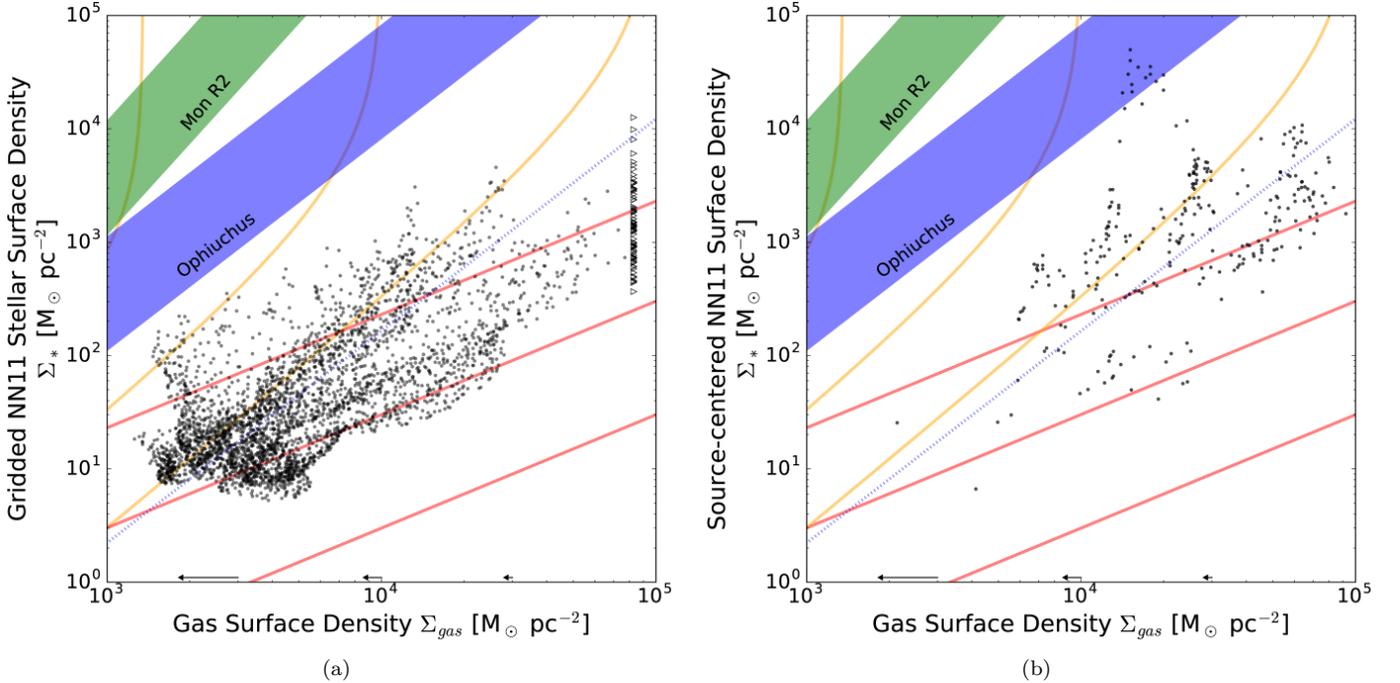
{Plots of the protostellar mass surface density vs the gas mass surface density
as derived from Herschel SED fitting (Section \ref{sec:colmaps}).  The stellar
mass surface densities are computed using the 11th
nearest-neighbor distance assuming that each star represents
a mass of 95 \msun, extrapolated assuming a uniform IMF.  (a) shows the
densities computed on a 0.25 pc grid, with column density lower limits
indicated where the Herschel data are saturated,
while (b) shows
the protostar-centric surface densities; no lower limits are included in this
figure because interpolated mass surface densities are used instead.  The
shaded regions
show the extrapolations of the relations derived by \citet{Gutermuth2011a}
for Ophiuchus (blue) and Mon R2 (green); their data cut off below a mass
surface density $\Sigma < 10^3$ \msun pc$^{-2}$. 
The blue dotted line shows the Ophiuchus relation scaled down by
$50\times$ to overlap with our data.
The thick orange lines show realizations of the \citet{Gutermuth2011a} $\alpha=2$ star
formation relation at times 0.01, 0.1, and 0.74 Myr, from bottom to top.
Similarly, the thick red lines show realizations of the $\alpha=1$ star
formation relation at the same ages.
The arrows along
the bottom show the effect of subtracting a uniform foreground column
density of $N(\hh)=5\ee{22}$ \persc (1100 \msun pc$^{-2}$).
}
{fig:stellarvsmasscolumn}{1}{0.5\textwidth}

Figure \ref{fig:stellarvsmasscolumn}  shows in orange three curves from the
\citet{Gutermuth2011a} $\alpha=2$ star formation relation, their Equation 7
(our Equation \ref{eqn:stellarsurfdens}),
with $k=10^{-4}$ pc$^{2}$ \msun$^{-1}$ Myr$^{-1}$ and $\alpha=2$, at times
$t=0.01$, 0.1, and 0.74 Myr.  Only the youngest curve, with age 0.01 Myr,
overlaps with our data.  The three red curves, which are essentially lines in
this figure, show the $\alpha=1$ relation with $k=0.1$ Myr$^{-1}$ at the same
three ages, and they achieve reasonable agreement with our data for the $t=0.74$
Myr line ($k=0.1$ \permyr implies the 50\% depletion time $t_{sf}=7$ Myr).  The
$\alpha=2$ star formation relation  is only consistent with our
data for times earlier than $t<0.1$ Myr.  This inconsistency is due to the very
fast depletion time for this form of star formation relation, which decreases
with gas surface density.  Indeed, the $\alpha=2$ star formation relation used
by \citet{Gutermuth2011a} is completely implausible for the gas surface density
regime we observe, as it implies that gas with an initial surface density of
$\Sigma_{gas}=10^4$ \msun \perspc would achieve a star formation efficiency
$\epsilon>1$ in $t<0.1$ Myr.  While our data are clearly incompatible with the
$\alpha=2$ relation, they are reasonably compatible with a linear $\alpha=1$
relation with the same normalization used by \citet{Gutermuth2011a}, i.e.,
$k=0.1$ Myr$^{-1}$.

Figure \ref{fig:stellarvsmasscolumn} also shows that the extrapolated relation
from the low-mass clouds exceeds our observations by at least $50\times$
(Ophiuchus) or closer to $10^3\times$ (Mon R2).   The discrepancy between our
observations and theirs indicates either that there is a systematic tendency to
overestimate $\Sigma_*$ at high $\Sigma_{gas}$ in the Spitzer observations,
which seems unlikely, or that there is a different star formation-gas surface
density relation in Sgr B2 and in local clouds.

\subsubsection{A critical evaluation of the discrepancies with Gutermuth et al}
\label{sec:gutermuthdiscrepancy}
While a linear relation $\Sigma_* \propto \Sigma_{gas}$ can approximately
account for both local clouds and Sgr B2 as a whole, we have not yet explained
why the extrapolation of the observed $\Sigma_* - \Sigma_{gas}$ relation from
local clouds does not match Sgr B2.  We evaluate several possibilities here:

\begin{itemize}
    \item \emph{Could we be missing an older generation?}
        \citet{Gutermuth2011a} were sensitive to, and included in their sample,
        an older generation of Class II sources, which we cannot detect.
        However, they typically found a Class II / Class I ratio of only
        $\approx4\times$ \citep{Gutermuth2009a} (and they found that this
        ratio \emph{decreased} at higher gas surface densities), so the
        discrepancy cannot be exclusively due to our insensitivity to older
        YSOs unless the star formation rate within Sgr B2 was an order of
        magnitude higher 1-5 Myr ago.  Such an enhanced SFR is implausible
        since such a large population of massive stars would still be alive and
        very easily detectable in our survey and previous VLA surveys. 

    \item \emph{Could we be overestimating the gas mass?}
        The surface densities we measure cannot be substantially incorrect.
        Even if we assume the maximum plausible foreground cloud surface
        density of $N(\hh)=5\ee{22}$ \persc, the measured gas surface densities
        only shift by a small fraction (at most $50\%$, but typically $<10\%$
        for the star-centered measurements; see the arrows in Figure
        \ref{fig:stellarvsmasscolumn}).
        If the dust opacity or dust-to-gas ratio were substantially wrong,
        e.g., if the dust-to-gas ratio is 10 instead of 100, some of
        our data would begin to overlap with the local cloud data.
        If we had overestimated the gas mass by the required amount to
        bring our data into agreement with the local clouds, the star
        formation efficiency would be close to 50\% (i.e., $M_*\sim M_{gas}$),
        which is unlikely given the many signs of youth observed.

    \item \emph{Could there be high multiplicity in our sample?} 
        A possible explanation is that each of the detected sources in our
        sample is a high-number multiple system, such that each 3 mm source
        represents $\approx5000$ \msun instead of $\approx100$ \msun.  The
        multiplicity of the Orion Source I system suggests this interpretation
        is qualitatively plausible, but the factor of 50 required to match the
        extrapolation of the \citet{Gutermuth2011a} data strains credibility.
        Additionally, the luminosity constraints from our observed data
        rule this possibility out unless the stellar IMF is bottom-heavy
        (see below for more IMF discussion).

    \item \emph{Could the sources be much more massive than we have inferred?}
        Another possibility is that each source we detect has a higher minimum
        mass than we have assumed, $M\gg8 \msun$, but again the required
        threshold is absurd, requiring each star to be $>100$ \msun to match
        the local cloud extrapolation.  Such massive stars are incompatible
        with the observed 3 mm luminosities for any plausible dust envelope
        or \hii region model (see Section \ref{sec:classification}).

    \item \emph{Could our sample be incomplete?}
        If our sample were incomplete by a factor of 100-1000, our results
        would match those extrapolated from Gutermuth et al.  While Section
        \ref{sec:contsources} concedes that the catalog may be incomplete, it
        is unlikely we are $<1\%$ complete, and the catalog is almost certainly
        complete to $>90\%$ for very massive and luminous sources ($L>10^5$
        \lsun, see Section \ref{sec:theyareprotostars}).  Additionally, if
        we were to include a factor of $100-1000\times$ more stellar mass, the 
        implied total stellar mass would be absurd, reaching $10^6-10^7$ \msun,
        exceeding the cloud mass.

    \item \emph{Could Sgr B2 consist of several Mon R2-like clouds stacked
        along the line of sight?}
        If there were $\sim50-100$ clouds of the same physical scale and
        surface density stacked along the line of sight, the data in Figure
        \ref{fig:stellarvsmasscolumn} would shift left, providing a possible
        explanation of the difference.  However, besides the extreme
        unlikeliness of having so many clouds along the line of sight, this
        explanation would require that the majority are non-star-forming, i.e.,
        they would have to be extremely young.  Also, the observations do not
        favor this scenario, as most of the star formation appears associated
        with a single velocity component in the \cyanoacetylene data (e.g., 
        Figure \ref{fig:coreson20cmandhc3n}, Appendix \ref{sec:hc3nfigures}).
        Finally, the elongation of the cloud on the sky hints that it is not
        multiple clouds, since they would have to all have similar elongations.

    \item \emph{Is the stellar IMF spatially nonuniform?}
        Our stellar mass surface density measurements are predicated on the
        assumption that each MYSO represents a fully-sampled initial mass
        function at the same location.  If there is any spatial non-uniformity
        in the IMF, e.g., if massive stars preferentially form at the bottoms
        of large potential wells (``primordial mass segregation''), the massive
        stars will have a different spatial distribution than the low-mass
        stars.  This effect would result in a higher measured stellar surface
        density at the highest gas surface densities and a lower measured
        stellar surface density at the lowest gas surface densities, i.e., it
        would result in a steeper slope in Figure
        \ref{fig:stellarvsmasscolumn}.  Therefore, unless there is inverse mass
        segregation, a spatially nonuniform IMF cannot explain our
        observations.

    \item \emph{Is the stellar IMF temporally nonuniform?}
        If high-mass stars form first, we would overestimate the stellar mass
        surface density.  However, if low-mass stars form first, we could
        underestimate the stellar mass surface density.  Given our survey's
        insensitivity to low-mass YSOs, the stellar mass surface density could
        be over an order of magnitude higher if it consists only of low-mass
        YSOs.  Such a dramatic time sequencing effect in star formation
        would have profound implications for star formation studies, implying
        that any or all clouds currently forming low-mass stars may eventually
        form higher-mass stars, so testing this possibility with
        high-sensitivity observations should be a priority.

    \item \emph{Is the \emph{local} star formation efficiency lower at a fixed surface
        density in the Galactic center?}
        The overall star formation rate in the Galactic center is lower than
        expected given predictions from local clouds.  Changing the
        normalization of the star formation relation, i.e., reducing the prefactor
        $c=0.3$ to $c=0.01$, where $c$ is the fraction of gas in a core that
        makes it onto a star (the local efficiency), would allow our results to
        be consistent.  However, there is no evidence for any difference in the
        star formation process in the Galactic center once a core has formed;
        most evidence currently points to inefficient \emph{core} formation in
        the CMZ.

    \item \emph{Could the high star-formation threshold in the CMZ explain the
        difference?}
        As noted in Sections \ref{sec:ladathreshold} and
        \ref{sec:otherthresholds} above, forming stars only begin to appear
        above a threshold significantly higher than in local neighborhood
        clouds.  A simplistic model in which star formation simply does not
        occur below a fixed column threshold does not explain the difference
        between our data and Gutermuth's, however, because the disagreement
        occurs at the high column densities in which we do observe star
        formation.  On the other hand, a higher \emph{volume} density threshold
        is plausible.  Such a threshold would imply a lower stellar density at
        a given surface density and would  permit variations in the stellar
        surface density depending on how much dense gas is present.

\end{itemize}

Of the items above, only the final, which suggests that a surface-density-based
star formation law is inviable, satisfactorily explains the discrepancy between
our data and the extrapolation from \citet{Gutermuth2011a}.

\subsection{Interpretation of a varying threshold for star formation}
\label{sec:turbsftheory}
In Sections \ref{sec:thresholds} and \ref{sec:gutermuth}, we concluded that a
varying star formation density threshold is likely to exist in the CMZ.  Other
authors have come to the same conclusion based on observations of G0.253+0.016
\citep{Rathborne2014a,Kruijssen2014c}.  Here, we briefly discuss what may drive
such a varying threshold.

\citet{Federrath2012a} summarized and reformulated a variety of
turbulence-based star formation theories.  These theories assume that the gas
density in a molecular cloud is approximately lognormally distributed, with the
distribution's shape parameters governed by turbulent parameters, the most
important being the mean Mach number of the cloud.  In the models, gas above
some threshold density $n_{crit}$ becomes gravitationally unstable and
collapses to form stars.  The three models
\citep{Krumholz2005a,Padoan2011a,Hennebelle2011a} have different threshold
criteria.  Most importantly, the \citet{Krumholz2005a} and \citet{Padoan2011a}
threshold densities rise with increasing Mach number
($n_{crit}\propto\mathcal{M}^2$), while the \citet{Hennebelle2011a} threshold
decreases with Mach number ($n_{crit}\propto \mathcal{M}^{-2}$).  Since our
observations imply the need for a \emph{higher} threshold density, and Galactic
center clouds are more turbulent (higher Mach number) than local clouds
\citep[e.g.,][]{Federrath2016a}, the \citet{Hennebelle2011a} model is
qualitatively inconsistent with our observations.

\section{Conclusions}
\label{sec:conclusions}
We have reported the detection of \ncores 3 mm point sources in the extended
Sgr B2 cloud and determined that the majority are high-mass protostellar
cores.  This survey represents the first large population of YSOs
detected in the Galactic center and the largest sample yet reported
of high-mass YSOs.

The large population of high-mass protostellar cores indicates that an extended
region spanning the entire Sgr B2 cloud, not just the well-known clusters N, M,
and S, is undergoing a burst of star formation.  More than half of the
currently forming generation of stars is not associated with any of the
clusters but is instead part of the extended burst.

Using Herschel, SCUBA, and SHARC data, we have observed a threshold for
high-mass star formation analogous to that inferred in local clouds by
\citet{Lada2010a}.  We find that there are no high-mass YSOs in gas below
$N(\hh)<10^{23}$ \persc at a resolution of $\approx10\arcsec$ (0.4 pc), and
half of the detected sources are found above $N(\hh)>10^{24}$ \persc.  However,
there is abundant material above $N(\hh)>10^{23}$ \persc that has no
associated YSOs, indicating that this threshold is a necessary but not
sufficient criterion for high-mass star formation.  These measurements imply
either the existence of a higher threshold for high-mass star formation than
for low-mass, as predicted by several theories \citep[e.g.][]{Krumholz2008a},
or a higher threshold for star formation in the Galactic center as compared to
local clouds \citep[e.g., as proposed by][]{Kruijssen2014c,Rathborne2014a}.
Deeper observations recovering the low-mass sources are required to distinguish
these possibilities.

Comparing the protostellar mass surface density to the gas mass surface density
revealed a correlation compatible with the slopes observed by
\citet{Gutermuth2011a}, but with an amplitude significantly inconsistent with
theirs.  A star formation relation of the form $\Sigma_* \propto
\Sigma_{gas}^\alpha$ with $\alpha=2$ favored by \citet{Gutermuth2011a} cannot
explain our observations, though an $\alpha=1$ (linear) relation is consistent
with our data, and the $\alpha=1$ relation implies an age $t\sim1$ Myr that is
consistent with the \citet{Kruijssen2015a} dynamical model age for the Sgr B2
cloud $t=0.74$ Myr.

The extrapolation of the surface density relations from local clouds in
\citet{Gutermuth2011a} does not agree with our data.  We explored a wide
variety of possible explanations for the difference, and concluded that the
most likely is that a surface density relation is incapable of explaining both
local and CMZ clouds.  Instead, a volume-density based model, in which the
volume density threshold is higher in the CMZ, may be viable.

The large detected population of high-mass YSOs implies a much larger
population of as-yet undetectable lower-mass YSOs.  Future ALMA and JWST
programs to probe this population would provide the data needed to directly
compare star formation thresholds in the most intensely star-forming cloud in
our Galaxy to those in nearby clouds.

\software{
The software used to make this version of the paper is available from github at
\url{https://github.com/keflavich/SgrB2_ALMA_3mm_Mosaic/} with hash \githash
(\gitdate).  The tools used include \texttt{spectral-cube}, \texttt{radio-beam}, and
\texttt{uvcombine} from the
\texttt{radio-astro-tools} package
(\url{https://github.com/radio-astro-tools/spectral-cube},
(\url{https://github.com/radio-astro-tools/radio-beam},
(\url{https://github.com/radio-astro-tools/uvcombine}, and
\url{radio-astro-tools.github.io}), \texttt{astropy}
\citep{Astropy-Collaboration2013a}, \texttt{astroquery}
(\url{astroquery.readthedocs.io}) and \texttt{CASA} \citep{McMullin2007a}.
}

\textit{Acknowledgements}
We thank the anonymous referee for a very constructive and helpful report.
The National Radio Astronomy Observatory is a facility of the National Science
Foundation operated under cooperative agreement by Associated Universities,
Inc.
This paper makes use of the following ALMA data: ADS/JAO.ALMA\#2013.1.00269.S.
ALMA is a partnership of ESO (representing its member states), NSF (USA) and
NINS (Japan), together with NRC (Canada), NSC and ASIAA (Taiwan), and KASI
(Republic of Korea), in cooperation with the Republic of Chile. The Joint ALMA
Observatory is operated by ESO, AUI/NRAO and NAOJ.
This work is partly supported by a grant from the National Science Foundation
(AST-1615311, De Pree).  JMDK gratefully acknowledges funding from the German
Research Foundation (DFG) in the form of an Emmy Noether Research Group (grant
number KR4801/1-1), from the European Research Council (ERC) under the European
Union's Horizon 2020 research and innovation programme via the ERC Starting
Grant MUSTANG (grant agreement number 714907), and from Sonderforschungsbereich
SFB 881 ``The Milky Way System'' (subproject P1) of the DFG.
RGM acknowledges support from UNAM-PAPIIT program IA102817.
JC acknowledges support for this work provided by the NSF through the Grote
Reber Fellowship Program administered by Associated Universities, Inc./National
Radio Astronomy Observatory.
ASM, PS and FM are partially supported by Deutsche Forschungsgemeinschaft
through grant SFB956 (subproject A6).
JEP acknowledges the financial support of the European Research Council (ERC;
project PALs 320620).

\appendix

\section{Single Dish Combination}
\label{sec:singledishcomb}
To measure the column density at a resolution similar to \citet{Lada2010a}, we
needed to use ground-based single-dish data with resolution $\sim10\arcsec$.
We combined these images with Herschel data, which recover all angular
scales, to fill in the missing `short spacings' from the ground-based data.

Specifically, we combine the SHARC 350 \um \citep{Dowell1999a} and 
SCUBA 450 \um \citep{Pierce-Price2000a,di-Francesco2008a} with Herschel 350 and
500 \um data \citep{Molinari2016a}, respectively.

Combining single-dish with `interferometer' data, or data that are otherwise
insensitive to large angular scales, is not a trivial process.  The standard
approach advocated by the ALMA project is to use the `feather' process, in
which two images are fourier-transformed, multiplied by a weighting function,
added together, and fourier transformed back to image space \citep[see
equations in \S 5.2 of][]{Stanimirovic2002a}.  This process is subject to
substantial uncertainties, particularly in the choice of the weighting
function.  

Two factors need to be specified for linear combination: the beam size of the
`single-dish', or total power, image, and the largest angular scale of the
`interferometer' or filtered image.  While the beam size is sometimes
well-known, for single dishes operating at the top of their usable frequency
range (e.g., the CSO at 350 \um or GBT at 3 mm), there are uncertainties in the
beam shape and area and there are often substantial sidelobes.  In
interferometric data, the largest angular scale is well-defined in the
originally sampled UV data, but is less well-defined in the final image because
different weighting factors change the recovered largest angular scale.  For
ground-based filtered data, the largest recoverable angular scale is difficult
to determine 
\citep[e.g.,][]{Ginsburg2013a,Chapin2013a}.

To assess the uncertainties in image combination, particularly on the
brightness distribution \citep[e.g.,][]{Ossenkopf-Okada2016a}, we have performed
a series of experiments combining the Herschel with the SCUBA data using
different weights applied to the SCUBA data.  As discussed in Section
\ref{sec:observations}, we empirically determined the scale factor required for
the best match between SCUBA and Herschel data was $3\times$, which is
 large but justifiable.  In the experiment shown in Figure
\ref{fig:feathercompare}, we show the images and resulting histograms when we
combine the Herschel data with the SCUBA data scaled by a range of factors from
$0.5\times$ to $10\times$.  The changes to the high end of the histogram are
dramatic, but the middle region containing most of the pixels (and most
relevant to the discussion of thresholds in the paper) is not substantially affected.
Additionally, we show the cumulative distribution function of core background
surface brightnesses (as in Figure \ref{fig:corebackgroundcdf}), showing again
that only the high end is affected. 

\Figure{f22}
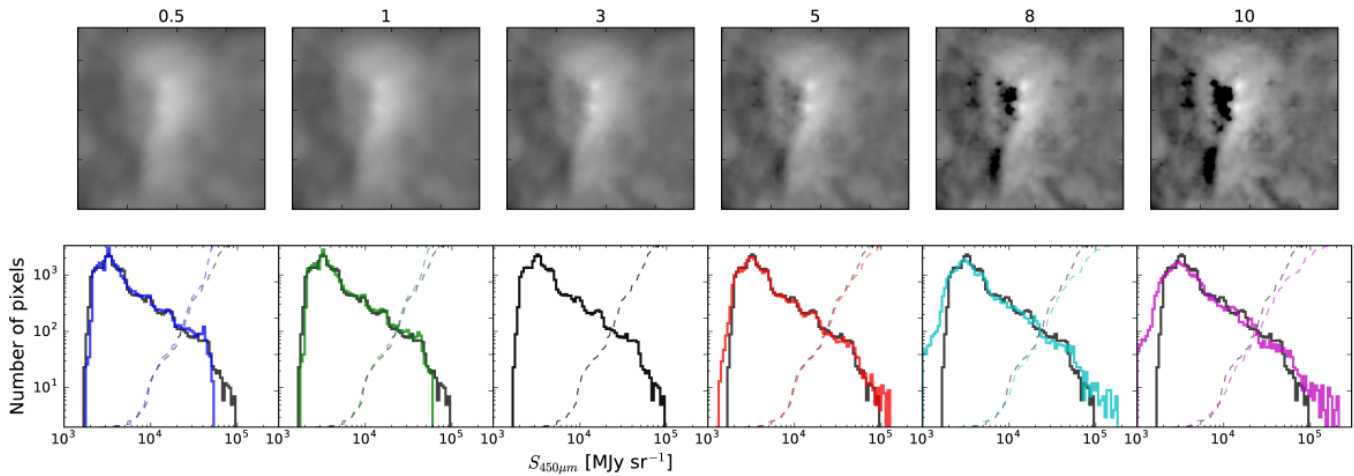
{A demonstration of the effects of using different calibration factors when
combining the SCUBA data with the Herschel data using the `feather' process.
The numbers above each panel show the scale factor applied to the SCUBA data
before fourier-combining it with the Herschel data.  The factor of 3 was used
in this paper and shows the most reasonable balance between the high-resolution
of the SCUBA data and the all-positive Herschel data.  In the lower panels, the
fiducial scale factor of 3 is shown in black in all panels.  The solid lines
show histograms of the images displayed in the top panels.  The dashed lines
show the cumulative distribution of the background surface brightnesses of the
point sources in this sample; they are similar to the distributions shown in
Figure \ref{fig:corebackgroundcdf}.}
{fig:feathercompare}{1}{\textwidth}

\section{Self-calibration}
\label{sec:selfcal}
We demonstrate the impact of self-calibration in this section.  The adopted approach
used three iterations of phase-only self-calibration followed by two iterations of
phase and amplitude self-calibration.  Each iteration involved slightly
different imaging parameters.  The final, deepest clean used a threshold mask
on the previous shallower clean. The script used to produce the final images is
available at
\url{https://github.com/keflavich/SgrB2_ALMA_3mm_Mosaic/blob/\githash/script_merge/selfcal_continuum_merge_7m.py}.
The effects are shown with a cutout centered on the most affected region around
Sgr B2 M in Figure \ref{fig:selfcalprogression}.

\Figure{f23}
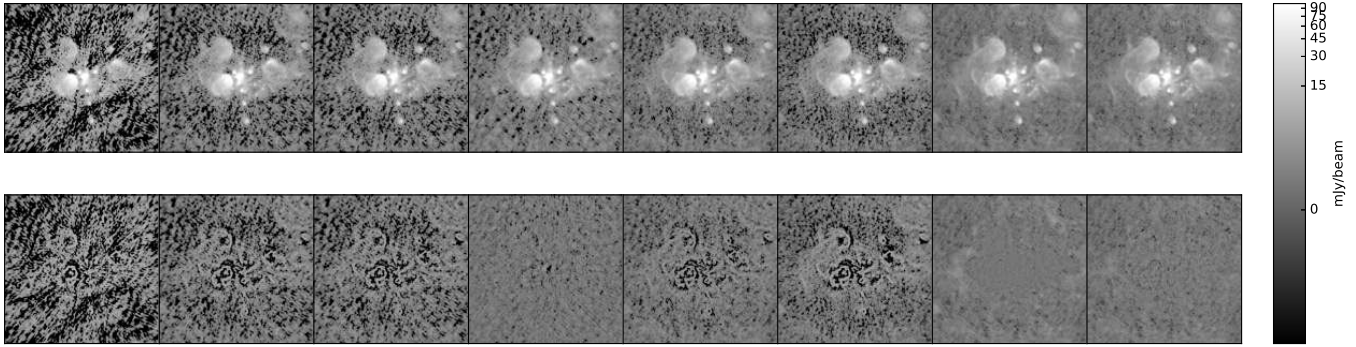
{Progression of the self-calibration iterations.  The images show, from left to
right, the initial image, one, two, and three iterations of phase-only self
calibration, two iterations of phase and amplitude self-calibration,  a
reimaging of the 5th iteration with a deeper 0.1 mJy threshold using a mask at
the 2.5 mJy level, and finally, a sixth iteration of phase and amplitude
self-cal cleaned to 0.1 mJy over a region thresholded at 1.5 mJy.  All imaging
was done using two Taylor terms and multiscale clean.  The second row shows the
corresponding residual images.}
{fig:selfcalprogression}{1}{\textwidth}

\section{Photometric Catalog}
\label{sec:catalog}
We include the full catalog in digital form
(\url{https://github.com/keflavich/SgrB2_ALMA_3mm_Mosaic/blob/master/tables/continuum_photometry_withSIMBAD_andclusters.ipac}).
Table~\ref{tab:photometry} shows
the brightest 35 sources; the rest are included in a digital-only catalog.
Sources are labeled based on an arbitrary source
number plus any pre-existing catalog name.  If a source is associated with a cluster,
it has an entry corresponding to that cluster in the \texttt{Cluster} column;
association is determined by checking whether a source is within a particular distance
of the cluster center as defined by \citet{Schmiedeke2016a}.  A source
\texttt{Classification} column is included, which states whether the source
is a strong or weak detection, whether it has an X-ray association, whether it
has a maser association, and its SIMBAD classification if it has one.
Measurements reported include the peak flux density $S_{\nu,max}$, the
corresponding brightness temperature $T_{B,max}$, the integrated flux density
within a beam (0.5\arcsec) radius, the background RMS flux level $\sigma_{bg}$
as an estimate of the local noise, the spectral index $\alpha$ and the error on
that $E(\alpha)$.  Mass and column density estimates are given for an assumed
temperature $T=40$ K ($M_{40K}$ and $N(\hh)_{40K}$).  For sources with
$T_{B,max}\gtrsim20$ K, these estimates are unlikely to be useful since the
assumed temperature is probably lower than the true temperature.
For sources with $T_{B,max}>40$ K, it is not possible to measure a mass
assuming $T=40$ K, so those entries are left empty.

\begin{table}[htp]
\caption{Continuum Source IDs and photometry}
\resizebox{\textwidth}{!}{
\begin{tabular}{llllllllllll}
\label{tab:photometry}
ID & Cluster & Classification & $S_{\nu,max}$ & $T_{B,max}$ & $S_{\nu,tot}$ & $\sigma_{bg}$ & $\alpha$ & $E(\alpha)$ & $M_{40K}$ & $N(\hh)_{40 K}$ & Coordinates \\
 &  &  &  &  &  &  &  &  &  &  &  \\
\hline
174 f3 & M & S\_\_W HII & 1600 & 860 & 2400 & 46 & 0.89 & 0.002 & - & - & 17:47:20.167 -28:23:04.809 \\
234 f4 & M & S\_\_W HII & 1100 & 570 & 900 & 23 & 0.83 & 0.001 & - & - & 17:47:20.214 -28:23:04.379 \\
176 f1 & M & S\_\_W HII & 920 & 480 & 1400 & 30 & 1.2 & 0.006 & - & - & 17:47:20.127 -28:23:04.082 \\
236 f10.303 & M & S\_\_W HII & 880 & 460 & 800 & 20 & 1.1 & 0.015 & - & - & 17:47:20.106 -28:23:03.729 \\
235 f2 & M & S\_\_W HII & 820 & 430 & 670 & 33 & 1.3 & 0.002 & - & - & 17:47:20.166 -28:23:03.714 \\
172 K2 & N & S\_\_W HII & 370 & 200 & 650 & 49 & 2.5 & 0.018 & - & - & 17:47:19.869 -28:22:18.466 \\
265 H & S & S\_\_W HII & 360 & 190 & 580 & 3.9 & 0.65 & 0.019 & - & - & 17:47:20.461 -28:23:45.404 \\
175 G & M & S\_\_W HII & 340 & 180 & 390 & 5.6 & 0.68 & 0.03 & - & - & 17:47:20.285 -28:23:03.162 \\
237 G10.44 & M & S\_\_W HII & 280 & 140 & 160 & 15 & 0.69 & 0.006 & - & - & 17:47:20.241 -28:23:03.387 \\
178 f10.37 & M & SX\_W HII & 200 & 100 & 270 & 18 & 1.5 & 0.039 & - & - & 17:47:20.178 -28:23:06 \\
171 K3 & N & S\_\_W HII & 180 & 97 & 280 & 25 & 1.4 & 0.023 & - & - & 17:47:19.895 -28:22:17.221 \\
177 B & M & S\_\_\_ HII & 150 & 77 & 240 & 3.9 & 0.47 & 0.011 & - & - & 17:47:19.918 -28:23:03.039 \\
241 f10.30 & M & S\_\_W HII & 140 & 73 & 120 & 15 & 1.4 & 0.05 & - & - & 17:47:20.106 -28:23:03.066 \\
179 f10.38 & M & S\_\_W HII & 130 & 66 & 180 & 9.3 & 1.6 & 0.013 & - & - & 17:47:20.193 -28:23:06.673 \\
180 E & M & S\_\_\_ HII & 130 & 66 & 190 & 4 & 0.38 & 0.014 & - & - & 17:47:20.108 -28:23:08.894 \\
173 K1 & N & S\_\_\_ HII & 92 & 48 & 150 & 4.4 & 0.58 & 0.034 & - & - & 17:47:19.78 -28:22:20.743 \\
170 & N & S\_\_W PartofCloud & 92 & 48 & 160 & 22 & 1.7 & 0.082 & - & - & 17:47:19.895 -28:22:13.621 \\
252 & N & S\_\_W denseCore & 82 & 43 & 160 & 16 & 1.9 & 0.078 & - & - & 17:47:19.862 -28:22:13.168 \\
225 f10.33b & M & SX\_W denseCore & 69 & 36 & 100 & 14 & 1.9 & 0.21 & 1200 & 3.6\ee{26} & 17:47:20.116 -28:23:06.374 \\
264 k4 & -- & S\_\_\_ HII & 65 & 34 & 140 & 3.5 & 0.57 & 0.034 & 1100 & 2.6\ee{26} & 17:47:19.997 -28:22:04.648 \\
96 Z10.24 & -- & S\_MW Maser & 64 & 33 & 75 & 1.5 & 0.68 & 0.37 & 1100 & 2.5\ee{26} & 17:47:20.039 -28:22:41.25 \\
181 D & M & S\_M\_ HII & 59 & 31 & 94 & 1.3 & 0.64 & 0.088 & 1000 & 2\ee{26} & 17:47:20.051 -28:23:12.91 \\
240 f10.44b & M & S\_\_W HII & 57 & 30 & 51 & 11 & 1.8 & 0.016 & 960 & 1.8\ee{26} & 17:47:20.252 -28:23:06.463 \\
233 f10.27b & M & S\_\_W HII & 50 & 26 & 78 & 18 & 2.3 & 0.18 & 840 & 1.4\ee{26} & 17:47:20.077 -28:23:05.383 \\
239 & M & S\_\_W denseCore & 45 & 24 & 46 & 8.6 & 2.3 & 0.091 & 760 & 1.1\ee{26} & 17:47:20.242 -28:23:07.222 \\
244 C & M & S\_\_\_ - & 36 & 19 & 67 & 0.49 & 0.47 & 0.081 & 600 & 7.8\ee{25} & 17:47:19.981 -28:23:18.437 \\
242 f10.318 & M & S\_\_W HII & 32 & 17 & 63 & 8.5 & 2.2 & 0.099 & 540 & 6.8\ee{25} & 17:47:20.129 -28:23:02.247 \\
92 I10.52 & M & S\_\_\_ HII & 32 & 17 & 45 & 5.3 & 0.63 & 0.061 & 530 & 6.6\ee{25} & 17:47:20.324 -28:23:08.2 \\
245 A2 & -- & S\_\_\_ HII & 24 & 13 & 32 & 2.1 & 0.54 & 0.025 & 410 & 4.8\ee{25} & 17:47:19.562 -28:22:55.916 \\
109 & N & S\_\_W - & 24 & 13 & 41 & 13 & 3.6 & 0.3 & 410 & 4.7\ee{25} & 17:47:19.901 -28:22:15.54 \\
87 B9.99 & M & S\_\_\_ HII & 23 & 12 & 37 & 1.9 & 0.89 & 0.042 & 390 & 4.4\ee{25} & 17:47:19.798 -28:23:06.942 \\
88 & M & S\_\_W - & 23 & 12 & 34 & 2.9 & 3.1 & 0.18 & 380 & 4.3\ee{25} & 17:47:19.617 -28:23:08.26 \\
151 B10.06 & M & S\_M\_ HII & 20 & 11 & 31 & 1.3 & 0.19 & 0.79 & 350 & 3.8\ee{25} & 17:47:19.86 -28:23:01.5 \\
98 & -- & S\_M\_ Maser & 18 & 9.5 & 29 & 0.36 & 3.2 & 1.1 & 300 & 3.3\ee{25} & 17:47:19.53 -28:22:32.55 \\
\hline
\end{tabular}
}\par
The Classification column consists of three letter codes as described in Section \ref{sec:classification}.  In column 1, \texttt{S} indicates a strong source, \texttt{W} indicates weak or low-confidence source. In column 2, an \texttt{X} indicates a match with the \citet{Muno2009a} Chandra X-ray source catalog, while an underscore indicates there was no match.  In column 3, \texttt{M} indicates a match with the, \citet{Caswell2010a} Methanol Multibeam Survey \methanol maser catalog, while an underscore indicates there was no match.  Finally, we include the SIMBAD \citep{Wenger2000a} source object type classification if one was found.  The full electronic version of this table is available at \url{https://github.com/keflavich/SgrB2_ALMA_3mm_Mosaic/blob/master/tables/continuum_photometry_withSIMBAD_andclusters.ipac} and will be made available via the journal at the time of publication.
\end{table}

\section{Additional figures showing \cyanoacetylene}
\label{sec:hc3nfigures}
The \cyanoacetylene line was discussed at various points in the paper.  Because
the data are extremely rich and complex,  we include some additional figures
showing the detailed
structure of the lines here.

\Figure{f24}
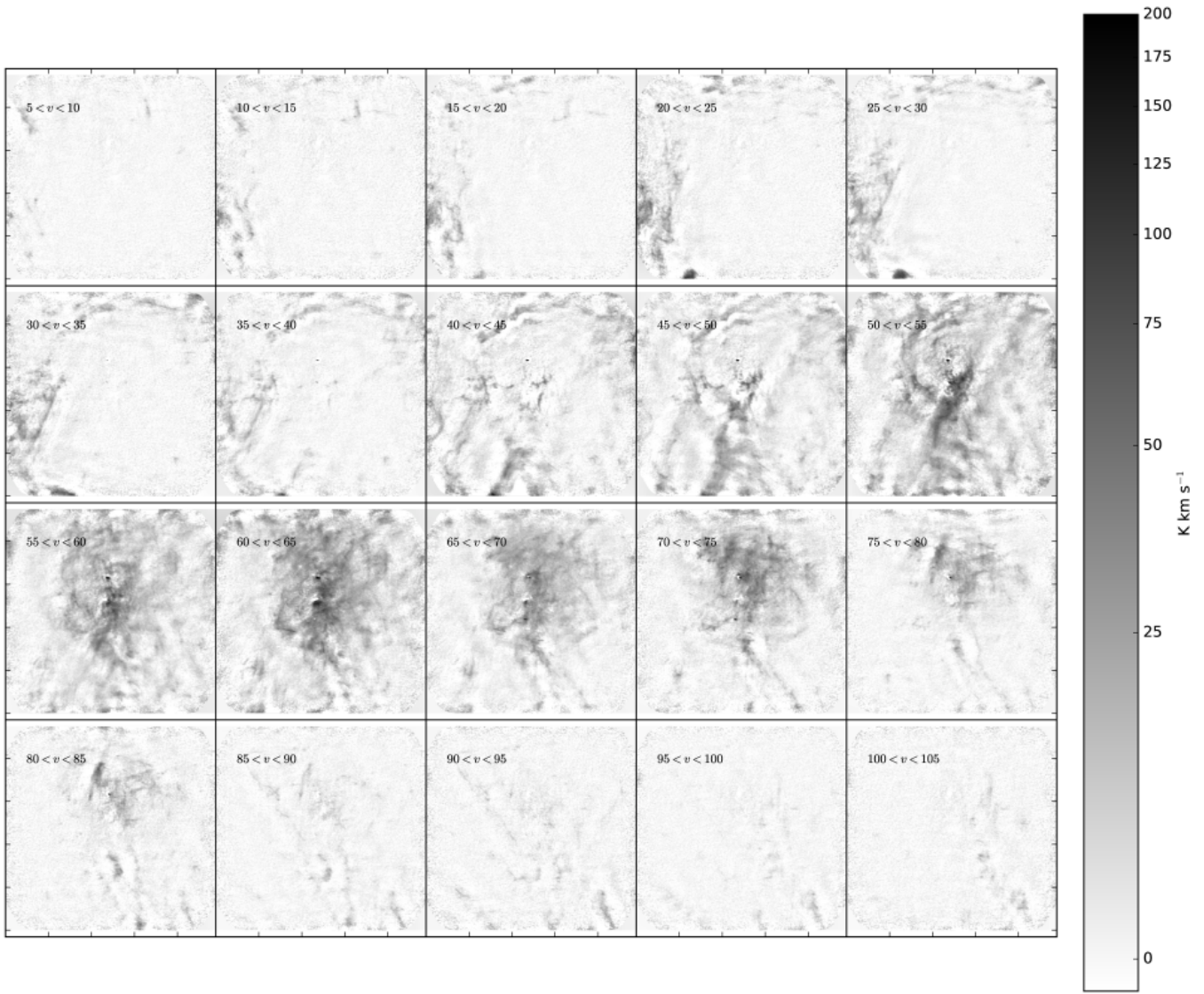
{Channel maps of the \cyanoacetylene J=10-9 line.  Each panel shows the integrated
intensity over a 5 \kms velocity range as indicated on the figures.
The data shown here are 
12m+7m images made excluding the long-baseline data sets to emphasize
large angular scales
combined with total power
data by feathering the images.
The `ridge' feature  discussed in the text is most evident in the 50-55 \kms
channel, and these images show that it is dominated by a single velocity
component.
}
{fig:hc3nchannelmaps}{1}{\textwidth}

\FigureTwo
{f25}
{f26}
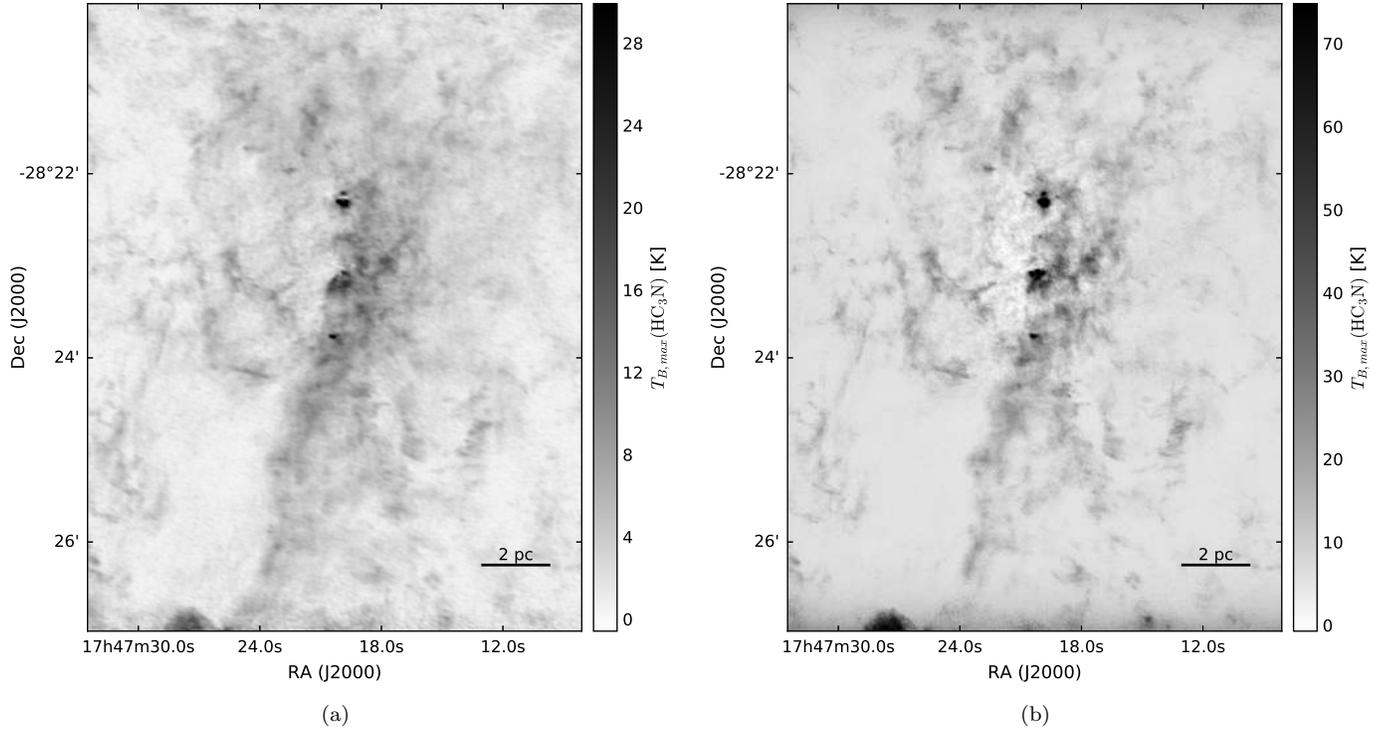
{Peak intensity maps of \cyanoacetylene J=10-9.
The left image shows the 12m short-baseline data combined with 7m and total
power data; by excluding the long-baseline data, the large angular scales are
emphasized.  The right image shows the robust 0.5-weighted 12m+7m data combined
with total power data;
it reaches a substantially higher peak intensity in the compact regions, but
the lower-intensity diffuse emission is relatively hidden.  In the right image,
the negative bowls seen near Sgr B2 M and N in this peak-intensity image
indicate that intermediate size scales were not well-recovered.  The bright
feature on the bottom-left of both images may be an imaging artifact.}
{fig:hc3npeakintensity}{1}{0.5\textwidth}

\section{Additional figure showing Sgr B2 M and N}
\label{sec:onept3cm}
We show the Sgr B2 M and N source identifications overlaid on VLA 1.3 cm
continuum \citep{De-Pree2014a} in Figure \ref{fig:MandNzoomsVLA}.  This figure
highlights the differences between the wavelengths and provides a visual
verification that our classification of sources as \hii regions is reasonable.

\Figure{f27}
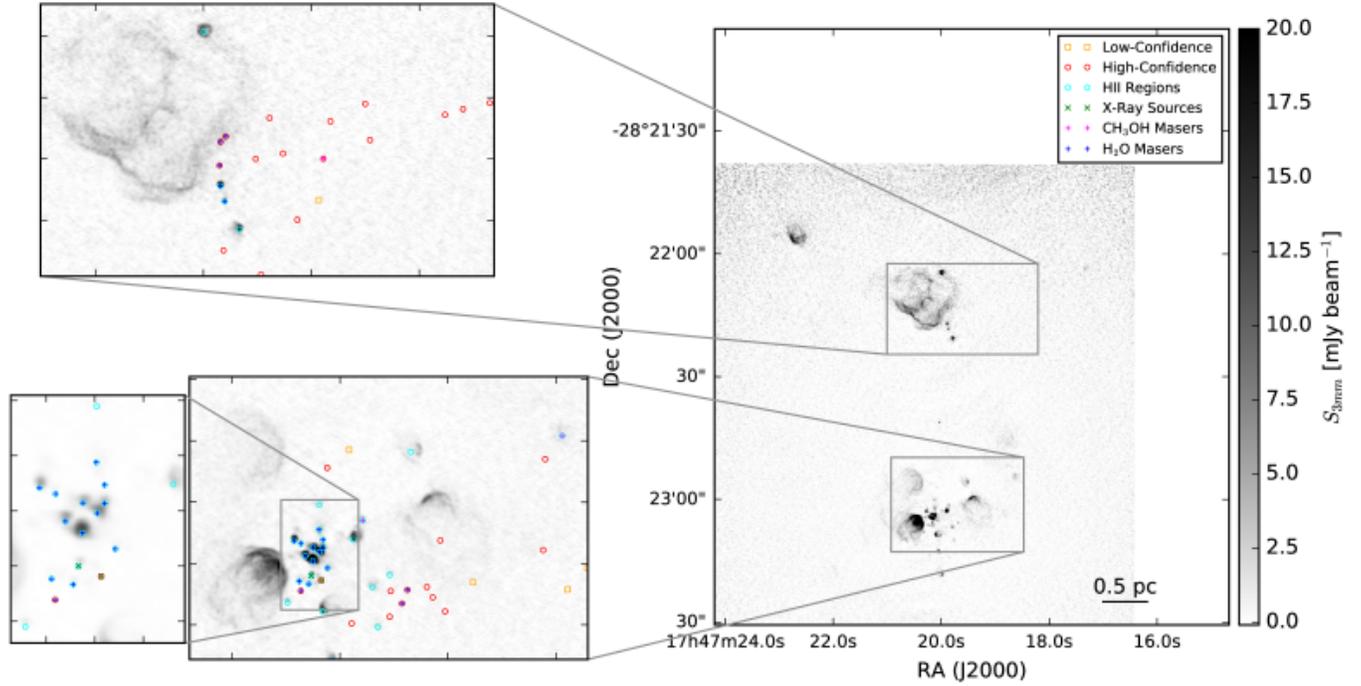
{A close-up of Sgr B2 M and N similar to Figure \ref{fig:MandNzooms}, but with
VLA 1.3 cm continuum \citep{De-Pree2014a} in the background instead of the ALMA 3 mm
continuum.  Many of the features that appear in the 3 mm image do not appear
in the 1.3 cm image and are likely to be from dust emission, but the poorer sensitivity
of the 1.3 cm data also suggests that some of these features are simply 
free-free emission undeteted at 1.3 cm.}
{fig:MandNzoomsVLA}{1}{\textwidth}

\section{Star-Gas Surface Density Figure with Lada et al 2017 relations}
\label{sec:ladasurfdensappendix}
We show in Figure \ref{fig:stellarvsmasscolumnlada} a version of Figure
\ref{fig:stellarvsmasscolumn} with the extrapolated relations from the Orion A,
Orion B, and California molecular clouds overlaid.  Similar to the comparison
to \citet{Gutermuth2011a} in Section \ref{sec:gutermuthcomparison}, the Lada et
al local clouds extrapolate to significantly higher stellar mass surface
densities than we observe in Sgr B2.

\FigureTwo
{f29}
{f30}
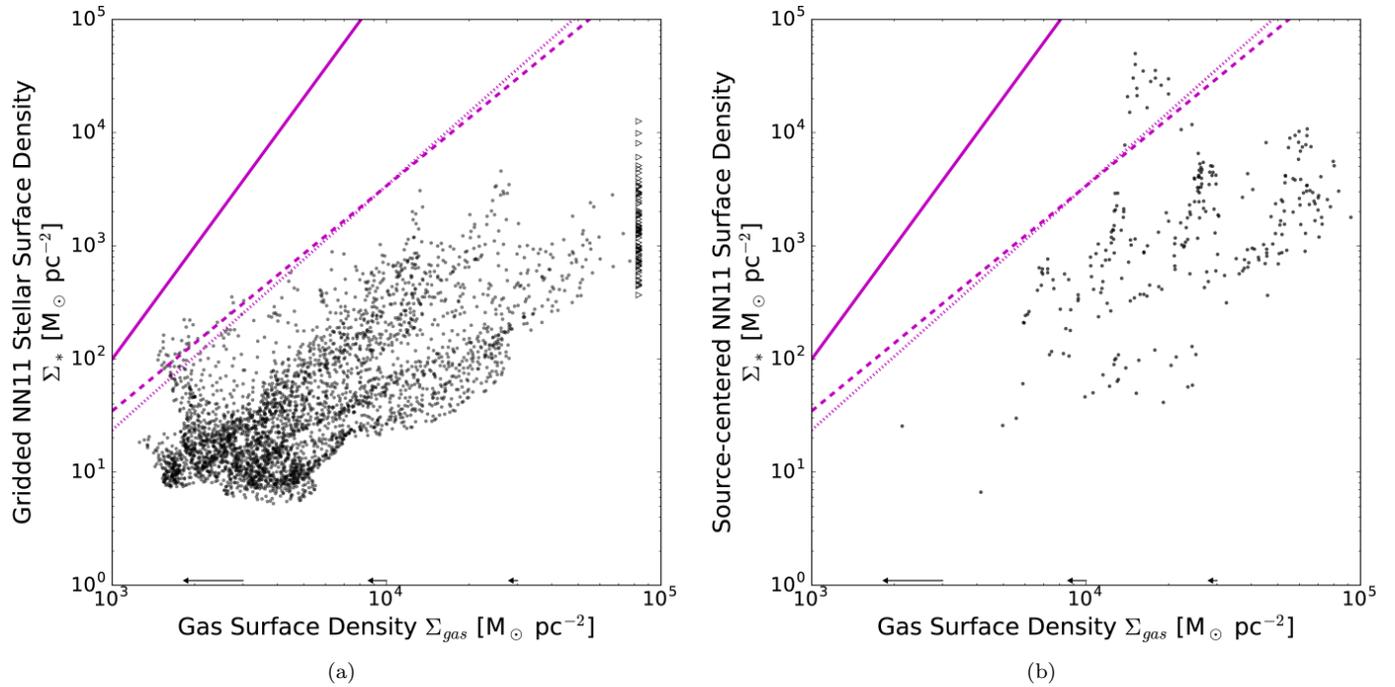
{The same plot shown in Figure \ref{fig:stellarvsmasscolumn},
but with the models and Gutermuth et al clouds removed and extrapolations
from the California (solid magenta), Orion A (dashed magenta), and Orion B
(dotted magenta) clouds overlaid.  As for the other local clouds, there is
\emph{no overlap} in the X-axis between our observations and theirs, so the
plotted relations are pure extrapolation.}
{fig:stellarvsmasscolumnlada}{1}{0.5\textwidth}

\end{document}